\documentclass[12pt,a4paper]{article} 
\usepackage{latexsym,graphicx,epsfig,psfrag,here,cite} 
\usepackage{bbm}
\newcommand{\ol}{\overline}

\newcommand{\bea}{\begin{eqnarray}} 
\newcommand{\eea}{\end{eqnarray}} 
\newcommand{\beq}{\begin{equation}} 
\newcommand{\eeq}{\end{equation}} 
\newcommand{\nn}{\nonumber} 
\newcommand{\nl}{\nonumber\\} 
 
\newcommand{\dg}{\dagger}

\newcommand{\ra}{\rightarrow}

\newcommand{\cL}{{\cal L}}

\newcommand{\ba}{\begin{array}{c}} 
\newcommand{\bat}{\begin{array}{cc}} 
\newcommand{\ea}{\end{array}} 
 
\newcommand{\mbf}{\mathbf}  
 
\newcommand{\Mvariable}{} 
 
\newcommand{\Mfunction}{} 
 
\newcommand{\qqo}{p_a\cdot p_b}  
\newcommand{\glan}{\begin{equation}} 
\newcommand{\glaus}{\end{equation}} 
\newcommand{\glanf}{\begin{eqnarray}} 
\newcommand{\glausf}{\end{eqnarray}} 
\newcommand{\glei}[1]{Eq. (\ref{#1})}

\newcommand{\raum}{\;\;\;\;\;} 
\newcommand{\ii}{\mbox{i}}

\newcommand{\ub}{\bar{u}} 
 
\newcommand{\gk}{\mbox{\Large (}} 
\newcommand{\gK}{\mbox{\Large )}} 
\newcommand{\sgk}{\mbox{\huge (}} 
\newcommand{\sgK}{\mbox{\huge )}}

\newcommand{\kk}{\langle} 
\newcommand{\KK}{\rangle} 
\newcommand{\Tr}{\,\mbox{Tr}\,} 
 
\newcommand{\del}{\Delta} 
 
\newcommand{\dmuu}{\partial_\mu} 
 
\newcommand{\dnuu}{\partial_\nu} 
 
\newcommand{\dx}{d^4x} 
\newcommand{\dy}{d^4y} 
\newcommand{\dz}{d^4z} 
\newcommand{\ddx}{d^dx}

\newcommand{\ddp}{d^dp}

\newcommand{\aaa}{_a} 
\newcommand{\bbb}{_b} 
\newcommand{\ppm}{p_{a_\mu}} 
\newcommand{\PPm}{p_{b_\mu}} 
\newcommand{\ppn}{p_{a_\nu}} 
\newcommand{\PPn}{p_{b_\nu}} 
\newcommand{\ppl}{p_{a_\lambda}} 
\newcommand{\PPl}{p_{b_\lambda}} 
 
\newcommand{\ddk}{\frac{d^dk}{(2\pi)^d}\,} 
\newcommand{\gmu}{\hat{\Gamma}^\mu} 
\newcommand{\gmuu}{\hat{\Gamma}_\mu\mbox{}} 
\newcommand{\gnu}{\hat{\Gamma}^\nu} 
\newcommand{\gnuu}{\hat{\Gamma}_\nu\mbox{}} 
\newcommand{\glambda}{\hat{\Gamma}^\lambda} 
\newcommand{\glambdau}{\hat{\Gamma}_\lambda\mbox{}} 
\newcommand{\grho}{\hat{\Gamma}^\rho} 
\newcommand{\grhou}{\hat{\Gamma}_\rho\mbox{}} 
\newcommand{\sig}{\bar{\sigma}} 
\newcommand{\dxmu}{\partial_{x_\mu}}

\newcommand{\dynu}{\partial_{y_\nu}} 
 
\newcommand{\dzmu}{\partial_{z_\mu}} 
 
\newcommand{\dzlam}{\partial_{z_\lambda}}

\newcommand{\A}{\mbox{}_{QP}} 
\newcommand{\B}{\mbox{}_{PR}} 
\newcommand{\C}{\mbox{}_{RQ}} 
\newcommand{\bk}{\mbox{}_{PQ}} 
\newcommand{\as}{\mbox{}_{QR}} 
\newcommand{\bs}{\mbox{}_{RP}} 
\newcommand{\cs}{\mbox{}_{PQ}} 
\def\slashchar#1{\setbox0=\hbox{$#1$}\dimen0=\wd0%
\setbox1=\hbox{/}\dimen1=\wd1%
\ifdim\dimen0>\dimen1%
\rlap{\hbox to 
\dimen0{\hfil/\hfil}}#1\else                                      
\rlap{\hbox to \dimen1{\hfil$#1$\hfil}}/\fi} 
\newcommand{\ve}{\varepsilon}  
\newcommand{\vp}{\varphi}

\normalsize  
\begin{document}  
\begin{titlepage}  
\begin{flushright} 
PSI-PR-05-06\\ 
UWThPh-2005-6\\  
July 2005\\  
\end{flushright}  
\vspace{2cm}  
\begin{center}  
{\Large \bf Generating Functional \\[15pt] 
for Strong and  Nonleptonic Weak Interactions$^*$} \\[40pt]
 
R. Unterdorfer$^{1}$ and G. Ecker$^{2}$ 
  
\vspace{1cm} 
${}^{1)}$ Paul Scherrer Institut, CH-5232 Villigen PSI, Switzerland 
 \\[10pt] 
  
${}^{2)}$ Institut f\"ur Theoretische Physik, Universit\"at  
Wien\\ Boltzmanngasse 5, A-1090 Vienna, Austria \\[10pt]  
\end{center}  
  
\vfill  
 
\begin{abstract}\noindent 
The generating functional for Green functions of quark currents is 
given in closed form to next-to-leading order in the low-energy 
expansion for chiral $SU(3)$, including one-loop amplitudes with up  
to three meson propagators. Matrix elements and form factors for  
strong and nonleptonic weak processes with at 
most six external states can be extracted from this functional by  
performing three-dimensional flavour traces. To implement this  
procedure, a Mathematica$^{\textrm{\scriptsize\copyright}}$ program  
is provided that evaluates amplitudes with at most six external  
mesons, photons (real or virtual) and virtual  
$W^\pm$ (semileptonic form factors). The program is illustrated  
with several examples that can be compared with existing calculations.  
 
\end{abstract} 
 
\vfill 
  
\noindent  
*~Work supported in part by HPRN-CT2002-00311 (EURIDICE)  
\end{titlepage}  
\newpage 
 
\addtocounter{page}{1}  
 
\section{Introduction} 
\renewcommand{\theequation}{\arabic{section}.\arabic{equation}} 
\setcounter{equation}{0} 
Chiral perturbation theory  
\cite{Weinberg:1978kz,Gasser:1983yg,Gasser:1984gg,Leutwyler:1993iq} is 
the effective field theory of the standard model at low energies. In 
the mesonic sector, the state of the art is next-to-next-to-leading 
order in the low-energy expansion for strong processes 
\cite{Bijnens:2004pk} and  
next-to-leading order for nonleptonic weak transitions.  
 
Although most one-loop amplitudes for processes of physical interest 
have already been calculated a compact closed form for such 
amplitudes would still be useful, both in principle and in practice. The 
generating functional for Green functions of quark currents is the 
relevant object for this purpose. The generating functional to 
next-to-leading order was already calculated in the classic papers of 
Gasser and Leutwyler \cite{Gasser:1983yg,Gasser:1984gg} for the   
strong interactions, including electromagnetic and semileptonic weak 
form factors, and by Kambor, Missimer and Wyler 
\cite{Kambor:1989tz} for nonleptonic weak processes. In both cases, 
the calculation was limited to processes where at most 
two meson propagators occur in one-loop amplitudes. 
 
More recently, the generating functional for chiral $SU(2)$ was 
obtained in closed form for transitions involving up to three 
propagators in one-loop amplitudes: at most three pions in  
Ref.~\cite{Unterdorfer:2002zg} and at most two pions and a photon in  
Ref.~\cite{Schweizer:2002ft}. The purpose of this 
article is to calculate the analogous generating functional with up to 
three meson propagators for chiral $SU(3)$ for both strong and 
nonleptonic weak transitions. The corresponding $SU(3)$ calculation 
with dynamical photons is in preparation \cite{schweizersu3}. 
 
Transitions involving at most three meson propagators in one-loop  
amplitudes comprise almost all processes of physical interest at  
next-to-leading order. More precisely, the complete strong and 
nonleptonic weak amplitudes and form 
factors to that order can be extracted from the generating functional 
for transitions of at most $O(\phi^6)$, i.e. with at most six external 
mesons, photons (counting as $O(\phi^2)$) and virtual $W^\pm$.  
Although most of such processes have already been calculated  
to next-to-leading order the generating functional provides a closed 
expression for such amplitudes with the same conventions and 
notation. As experience has shown, such a presentation greatly 
facilitates comparison with previous work and eventual detection of 
misprints and other errors. In addition to reproducing previous 
results in a simple and straightforward manner, 
several nonleptonic transitions have been calculated 
for the first time \cite{Isidori:2004rb} with the generating 
functional presented here. 
 
Although the extraction of amplitudes or form factors from the 
generating functional 
boils down to performing three-dimensional flavour traces such a 
procedure may turn out to be quite time consuming, especially for 
transitions with three-propagator loop contributions. To 
facilitate this work, we provide the  
Mathematica$^{\textrm{\scriptsize \copyright}}$  \cite{wolfram} program 
Ampcalculator, written by one of us (R.U.), that evaluates the 
necessary traces upon input of at most six external states (mesons, 
photons, virtual $W^\pm$) with their corresponding momenta. Use 
of the program is straightforward and does not require detailed 
knowledge of Mathematica$^{\textrm{\scriptsize \copyright}}$.  
  
The generating functional is introduced in Sec.~\ref{sec:Z} and its 
low-energy expansion is discussed. The generating functional of 
next-to-leading order is treated in Sec.~\ref{sec:Z1loop} for the 
strong interactions. We extend the work of Gasser and Leutwyler 
\cite{Gasser:1983yg,Gasser:1984gg} by calculating the explicit form of 
the one-loop functional for up to three meson propagators and to at 
most $O(\phi^6)$. The renormalization of the generating functional is 
briefly reviewed. In Sec.~\ref{sec:weak} the nonleptonic weak 
interactions are included. The additional terms in the generating 
functional linear in the weak coupling constants $G_8$, $G_{27}$ are 
calculated  under the same 
conditions as for the strong part. In Sec.~\ref{sec:Math} we discuss 
the Mathematica$^{\textrm{\scriptsize \copyright}}$  program 
Ampcalculator for evaluating amplitudes and form factors for a given 
set of external states to next-to-leading order. The program is 
exemplified for the decays $K^+ \to \pi^0 l^+ \nu_l ~(K^+_{l3})$,  
$K^- \to \pi^- \pi^0$, $K^0 \to \pi^0 \gamma \gamma$ 
and the results are compared with those in the 
literature. Our conclusions are summarized in 
Sec.~\ref{sec:concs}. Details about the input for the program 
Ampcalculator, one-loop integrals and constituent functions and the 
strong and weak Lagrangians of $O(p^4)$ are collected in three 
appendices.

\section{Generating functional of Green functions} 
\label{sec:Z} 
\renewcommand{\theequation}{\arabic{section}.\arabic{equation}} 
\setcounter{equation}{0} 
The generating functional for Green functions of quark currents is 
defined in terms of the vacuum transition amplitude in the presence of 
external fields, 
\begin{equation}  
e^{\displaystyle  i Z[v,a,s,p]} = <0~{\rm out}|0~{\rm in}>_{v,a,s,p}~, 
\label{eq:Zdef}  
\end{equation} 
associated with the Lagrangian \cite{Gasser:1983yg,Gasser:1984gg} 
\begin{equation}  
\cL = \cL^0_{\rm QCD} + \bar q \gamma^\mu(v_\mu + a_\mu \gamma_5)q - 
\bar q (s - ip \gamma_5)q ~,\label{eq:QCD} 
\end{equation} 
where $\cL^0_{\rm QCD}$ is the QCD Lagrangian with three massless 
quarks and $v_\mu, a_\mu,s,p$ are three-dimensional hermitian matrix 
fields.  
 
Green functions of quark currents are obtained by functional 
differentiation of $Z[v,a,s,p]$ in the usual way. To implement 
explicit chiral symmetry breaking, the external scalar 
field $s$ is set equal to the quark mass matrix 
${\cal M}=\mbox{diag}(m_u, m_d, m_s)$ at the end of the calculation. 
The external spin-1 fields generate electromagnetic and semileptonic 
form factors with the assignments 
\glan r^\mu \doteq v^\mu+a^\mu=-e\,Q A^\mu ~, 
\label{eq:extern} 
\glaus 
$$ 
l^\mu \doteq v^\mu-a^\mu=-e\,Q A^\mu-\frac{e}{\sqrt{2}\sin\theta_W} 
(W^{\mu^+}T_++\mbox{h.c.})   
$$ 
for external photons ($A_\mu$) and $W$ bosons ($W_{\mu}^\pm$). 
$\theta_W$ is the weak mixing angle, $Q$ is the quark charge matrix and 
$T_+$ contains the relevant elements of the Cabibbo-Kobayashi-Maskawa matrix: 
\glan Q=\left( \begin{array}{ccc}\frac{2}{3} & 0 & 0 \\ 0 & 
  -\frac{1}{3} & 0 \\  
0 & 0 & -\frac{1}{3} \\ \end{array} 
\right) ,\raum T_+=\left( \begin{array}{ccc} 0 & V_{ud} & V_{us} \\  
0 & 0 & 0 \\ 0 & 0 & 0 \\ \end{array} 
\right) ~. 
\glaus 
 
At the hadronic level, the generating functional may be calculated in
terms of an effective Lagrangian of pseudoscalar mesons and external 
fields \cite{Leutwyler:1993iq}, 
\begin{equation}  
e^{\displaystyle  i Z[v,a,s,p]} =  
\int [dU(\vp)] e^{\displaystyle  i \int d^4x\, {\cL}_{\rm eff}\,(U, v, 
a, s, p)}  ~.\label{eq:master} 
\end{equation} 
The mesonic effective chiral Lagrangian will be needed to 
next-to-leading order here, including both strong and nonleptonic weak 
parts: 
\begin{equation}  
{\cL}_{\rm eff}\,(U, v, a, s, p) = \cL_2 + \cL_4 + \dots \label{eq:Leff} 
\end{equation} 
The strong chiral Lagrangian takes the well-known form 
\cite{Gasser:1984gg}  
\begin{equation}  
\cL_2^S = \frac{F^2}{4} \langle D_\mu U D^\mu U^\dg + \chi U^\dg + 
\chi^\dg U \rangle ~,\qquad 
\chi = 2B(s + ip) \label{eq:L2S} 
\end{equation} 
in terms of the  pion decay constant and the quark condensate  
in the chiral limit: 
\beq 
F_\pi =  F[1 + O(m_q)] = 92.4 \,\mbox{MeV}  
\eeq 
$$\langle 0|\bar u u |0\rangle = - F^2 B[1 + O(m_q)] ~. 
$$
$\langle \dots \rangle$ denotes the 3-dimensional flavour trace. 
The strong chiral Lagrangian $\cL_4^S$ of $O(p^4)$ 
\cite{Gasser:1984gg} is reproduced in App.~\ref{app:lagr4} and the 
nonleptonic weak Lagrangian will be discussed in Sec.~\ref{sec:weak}. 
 
The chiral expansion of $\cL_{\rm eff}$ induces a corresponding expansion 
for the generating functional: 
\begin{equation}  
Z = Z_2 + Z_4 + \dots 
\end{equation}  
At lowest order, the functional $Z_2$  is equal to the classical action 
\begin{equation}  
Z_2[v,a,s,p] = \int d^4x\, \cL_2 (\bar{U},v,a,s,p) \label{eq:class} 
\end{equation}  
where $\bar{U} = \bar{U}[v,a,s,p]$ is to be understood as a functional of 
the external fields via the equation of motion (EOM) for $\cL_2$  
(given here for the strong interactions only)~: 
\begin{equation}  
\Box U U^\dg - U \Box U^\dg = \chi U^\dg - U \chi^\dg - 
{1 \over 3}\langle\chi U^\dg - U \chi^\dg\rangle \mathbbm{1}~. 
\label{eq:EOM} 
\end{equation}  
Mesonic amplitudes at lowest order can therefore be read off directly  
from the lowest-order Lagrangian (\ref{eq:L2S}) and the corresponding 
nonleptonic weak Lagrangian $\cL_2^W$ in (\ref{eq:G827}) by  
using an explicit parametrization of the matrix-valued meson field 
$U(\vp)$, e.g., the exponential parametrization  
\begin{equation}  
U(\vp)=\exp{(i\lambda_a \vp^a/F)} ~,\qquad 
{1 \over \sqrt{2}}\lambda_a \vp^a = \left(  
\begin{array}{ccc} 
\displaystyle\frac{\pi^0}{\sqrt{2}} + 
\displaystyle\frac{\eta_8}{\sqrt{6}} & \pi^+ &  K^+ \\ 
\pi^- & -\displaystyle\frac{\pi^0}{\sqrt{2}} +  
\displaystyle\frac{\eta_8}{\sqrt{6}} &  K^0 \\ 
K^- & \ol{K^0} & - \displaystyle\frac{2 \eta_8}{\sqrt{6}}  
\end{array}  \right)~. \label{eq:Uphi} 
\end{equation}  
The matrix field (\ref{eq:Uphi}) also defines our sign conventions for 
mesonic amplitudes generated by the Mathematica$^{\textrm{\scriptsize 
  \copyright}}$ program Ampcalculator to be discussed in 
Sec.~\ref{sec:Math}.

\section{Generating functional of $\mbf{O(p^4)}$} 
\label{sec:Z1loop} 
\renewcommand{\theequation}{\arabic{section}.\arabic{equation}} 
\setcounter{equation}{0} 
The generating functional at next-to-leading order consists of three 
parts: 
\begin{equation} 
Z_4 = Z_4^{\rm tree} + Z_4^{L=1} + Z_{\rm WZW}~. 
\label{eq:Z4} 
\end{equation} 
The tree-level part $Z_4^{\rm tree}$ is given by the action for the 
next-to-leading Lagrangian $\cL_4(U,v,a,s,p)$ to be taken again at the 
classical solution $\bar{U}$ satisfying the EOM (\ref{eq:EOM}). The 
Wess-Zumino-Witten functional $Z_{\rm WZW}$ 
\cite{Wess:1971yu} accounts for the chiral anomaly. We 
will not reproduce its explicit form here but it will be implemented 
in the Mathematica$^{\textrm{\scriptsize \copyright}}$ program  
Ampcalculator (cf. Sec.~\ref{sec:Math}).  
 
The one-loop functional $Z_4^{L=1}$ is calculated in the standard way. 
We repeat the main steps here in order to define our notation. The 
matrix field $U(\vp)$ is expanded around the classical solution 
$\bar{U}=\ub^2$ in terms of a traceless, hermitian 
fluctuation matrix $\xi$: 
\glan U=u^2=\ub(\mathbbm{1}+i\xi-\frac{1}{2}\xi^2+\dots)\ub ~. 
\label{eq:entu} 
\glaus 
Working with dimensional regularization in $d$ dimensions, one obtains 
\begin{equation}  
e^{\displaystyle  i Z_4^{L=1}}= 
\int\!d\mu[\xi]\;\mbox{e}^{\displaystyle 
  -\displaystyle\frac{i}{2}\,F^2  
\displaystyle\int \!\ddx\, \xi^a(x){\cal D}^{ab}(x)\xi^b(x)} ~, 
\qquad \xi={1 \over \sqrt{2}}\lambda_a \xi^a ~. 
\label{eq:ZL1} 
\end{equation}  
The measure $d\mu[\xi]$ is defined in such a way that $Z_4^{L=1}$ 
vanishes when the external fields $v$, $a$ and $p$ are set to zero and  
$s$ is set equal to the quark mass matrix. In the same limit, the 
differential operator ${\cal D}^{ab}$ turns into the Klein-Gordon 
operator ${\cal D}^{ab}_0$ with the appropriate meson masses to 
leading order in the chiral expansion\footnote{Since we work in the  
isospin limit throughout this 
  paper the differential operator ${\cal D}_0$ is diagonal.}. The 
one-loop functional can then be written in the form 
\begin{equation}  
Z_4^{L=1}=\displaystyle\frac{i}{2}\,\mbox{ln}\,\mbox{Det} 
\displaystyle\frac{\cal D}{{\cal D}_0} 
=\displaystyle\frac{i}{2}\,\mbox{Tr}\,\mbox{ln} 
\frac{\cal D}{{\cal D}_0} ~ . 
\label{eq:Zln} 
\end{equation}  
Splitting the differential operator ${\cal D}$ into the Klein-Gordon 
operator ${\cal D}_0$ and a remainder $\delta$, one gets 
\glan Z_4^{L=1}=\displaystyle\frac{i}{2}\,\mbox{Tr}\,\mbox{ln} 
\displaystyle\frac{{\cal D}_0+\delta}{{\cal 
    D}_0}=\displaystyle\frac{i}{2}\,\mbox{Tr}\,\mbox{ln}\, 
(\mathbbm{1}\,+\delta {\cal D}_0^{-1})=\frac{i}{2}\,
\mbox{Tr}\,\mbox{ln}\, 
(\mathbbm{1}\,-\delta\Delta) ~. 
\label{eq:ZTr} 
\end{equation}  
The diagonal matrix $\Delta$ is the inverse of $-{\cal D}_0$ and  
it contains the Feynman propagators of pseudoscalar mesons:  
\begin{equation}  
\Delta_{ab}(x-y)=\delta_{ab} 
\displaystyle\int\displaystyle\frac{\ddp}{(2\pi)^4} 
\displaystyle\frac{\mbox{e}^{-i p\cdot (x-y)}}{p^2-M_a^2 + i\epsilon}~. 
\end{equation}  
For the strong interactions $\delta$ is of the form \cite{Gasser:1984gg} 
\begin{equation}  
\delta_S = \{\gmu, \partial_\mu\} + \gmu\hat{\Gamma}_\mu 
+\bar{\sigma}   
\end{equation}  
with 
$$ 
\hat{\Gamma}^\mu_{ab} = 
-\frac{1}{2}\kk[\lambda_a, \lambda_b]\Gamma^\mu\KK 
~,$$ 
$$  
\bar{\sigma}_{ab} = \hat{\sigma}_{ab}-{M^2_a}\delta_{ab} 
~,$$ 
\glan \hat{\sigma}_{ab} =  
\frac{1}{2}\kk[\lambda_a, y_\mu][\lambda_b, y^\mu]\KK 
+\frac{1}{4}\kk\{\lambda_a, \lambda_b\}\sigma\KK 
~,\label{eq:sigma} \glaus 
$$ 
y_\mu = \frac{1}{2}\ub^\dag D_\mu \bar{U} \ub^\dag= 
-\frac{1}{2}\ub D_\mu \bar{U}^\dag \ub 
~,$$ 
$$ 
\Gamma_\mu = \frac{1}{2}[\ub^\dag,\partial_\mu \ub] 
-\frac{1}{2}i\ub^\dag r_\mu \ub - \frac{1}{2}i\ub l_\mu \ub^\dag 
~,$$ 
$$ 
\sigma = \frac{1}{2} (\ub \chi^\dag \ub + \ub^\dag \chi \ub^\dag) 
~.$$ 
To include one-loop contributions with up to three meson propagators, 
the one-loop functional (\ref{eq:ZTr}) must be expanded to third 
order in $\delta$: 
\glan Z_4^{L=1}=-\displaystyle\frac{i}{2}\,\Tr(\delta\Delta) 
-\displaystyle\frac{i}{4}\,\Tr(\delta\Delta\delta\Delta) 
-\displaystyle\frac{i}{6}\,\Tr(\delta\Delta\delta\Delta\delta\Delta)+  
\dots ~ . \label{eq:Zexpand} 
\glaus 
The functional (\ref{eq:Zexpand}) suffices for transitions up to 
$O(\phi^6)$, i.e. with at most six external mesons 
where an external photon counts as $O(\phi^2)$. We will therefore 
limit the further discussion to processes of at most $O(\phi^6)$. 
Since the external fields $v_\mu$ and $s$ couple to at least 
two pseudoscalar meson fields whereas $a_\mu$ and $p$  
couple to at least one field, the functional needs to be expanded 
to third order in $v_\mu$ and $s$ and to 
sixth order in $a_\mu$ and $p$. Keeping in mind that the matrices 
$\gmu$ and $\bar{\sigma}$ are of $O(\phi^2)$, the strong one-loop 
functional assumes the form 
\glanf  Z_4^{L=1}= & - \displaystyle\frac{i}{2} & 
\Tr\gk\sig(x)\del(0)+\gmu(x)\gmuu(x)\del(0)\gK+ 
\nonumber \\ 
& -\displaystyle\frac{i}{4} & \Tr\gk\{ \gmu(x), \dxmu\}\del(x-y) 
\{ \gnu(y), \dynu\}\del(y-x)+ 
\nonumber \\ 
& & +2\{ \gmu(x), \dxmu\} \del(x-y)\gnu(y)\gnuu(y)\del(y-x)+ 
\nonumber \\ 
& & {+2\{ \gmu(x), \dxmu\} \del(x-y)\sig(y)\del(y-x)}_ 
{}+ 
\nonumber \\ 
& & +2\gmu(x)\gmuu(x)\del(x-y)\sig(y)\del(y-x)+
\sig(x)\del(x-y)\sig(y)\del(y-x)\gK+ 
\nonumber \\ 
& -\displaystyle\frac{i}{6} & \Tr\gk\{ \gmu(x), \dxmu\}\del(x-y) 
\{ \gnu(y), \dynu\}\del(y-z)\{ \glambda(z), \dzlam\}\del(z-x)+ 
\nonumber \\ 
& & +3\{ \gmu(x), \dxmu\}\del(x-y) 
\{ \gnu(y), \dynu\}\del(y-z)\sig(z)\del(z-x)+ 
\nonumber \\ 
& & +3\sig(x)\del(x-y) 
\sig(y)\del(y-z)\{ \gmu(z), \dzmu\}\del(z-x)+ 
\nonumber \\ 
& & +\sig(x)\del(x-y)\sig(y)\del(y-z)\sig(z)\del(z-x)\gK + \dots 
\nonumber \\ 
= & - \displaystyle\frac{i}{2} & \Tr\gk\sig(x)\del(0)+ 
\gmu(x)\gmuu(x)\del(0)\gK+ 
\nonumber \\ 
& -\displaystyle\frac{i}{4} &  
\Tr\gk\gmu(x)\dxmu\del(x-y)\gnu(y)\dynu\del(y-x)+ \nonumber \\ 
& & +\gmu(x)\dynu\del(x-y)\gnu(y)\dxmu\del(y-x)+ 
\nonumber \\ 
& & -\gmu(x)\dxmu\dynu\del(x-y)\gnu(y)\del(y-x)+ 
\nonumber \\ 
& & -\gmu(x)\del(x-y)\gnu(y)\dxmu\dynu\del(y-x)+ 
\nonumber \\ 
& & +2\gmu(x)\dxmu\del(x-y)\gnu(y)\gnuu(y)\del(y-x)+ 
\nonumber \\ 
& & -2\gmu(x)\del(x-y)\gnu(y)\gnuu(y)\dxmu\del(y-x)+ 
\nonumber \\ 
& & +2\gmu(x)\dxmu\del(x-y)\sig(y)\del(y-x)+ 
\nonumber \\ 
& & -2\gmu(x)\del(x-y)\sig(y)\dxmu\del(y-x)+ 
\nonumber \\ 
& & +2\gmu(x)\gmuu(x)\del(x-y)\sig(y)\del(y-x)+ 
\nonumber \\ 
& & +\sig(x)\del(x-y)\sig(y)\del(y-z)\gK+ 
\nonumber \\ 
& -\displaystyle\frac{i}{6} &  
\Tr\gk \gmu(x)\dxmu\del(x-y)\gnu(y)\dynu\del(y-z) 
\glambda(z)\dzlam\del(z-x)+ 
\nonumber \\ 
& & -\gmu(x)\dynu\del(x-y)\gnu(y)\dzlam\del(y-z)\glambda(z)\dxmu\del(z-x)+ 
\nonumber \\ 
& & +\gmu(x)\dynu\del(x-y)\gnu(y)\del(y-z)\glambda(z)\dxmu\dzlam\del(z-x)+ 
\nonumber \\ 
& & +\gmu(x)\dxmu\dynu\del(x-y)\gnu(y)\dzlam\del(y-z)\glambda(z)\del(z-x)+ 
\nonumber \\ 
& & +\gmu(x)\del(x-y)\gnu(y)\dynu\dzlam\del(y-z)\glambda(z)\dxmu\del(z-x)+ 
\nonumber \\ 
& & -\gmu(x)\del(x-y)\gnu(y)\dynu\del(y-z)\glambda(z)\dxmu\dzlam\del(z-x)+ 
\nonumber \\ 
& & -\gmu(x)\dxmu\dynu\del(x-y)\gnu(y)\del(y-z)\glambda(z)\dzlam\del(z-x)+ 
\nonumber \\ 
& & -\gmu(x)\dxmu\del(x-y)\gnu(y)\dynu\dzlam\del(y-z)\glambda(z)\del(z-x)+ 
\nonumber \\ 
& & +3\gmu(x)\dxmu\del(x-y)\gnu(y)\dynu\del(y-z)\sig(z)\del(z-x)+ 
\nonumber \\ 
& & +3\gmu(x)\dynu\del(x-y)\gnu(y)\del(y-z)\sig(z)\dxmu\del(z-x)+ 
\nonumber \\ 
& & -3\gmu(x)\del(x-y)\gnu(y)\dynu\del(y-z)\sig(z)\dxmu\del(z-x)+ 
\nonumber \\ 
& & -3\gmu(x)\dxmu\dynu\del(x-y)\gnu(y)\del(y-z)\sig(z)\del(z-x)+ 
\nonumber \\ 
& & +3\sig(x)\del(x-y)\sig(y)\del(y-z)\gmu(z)\dzmu\del(z-x)+ 
\nonumber \\ 
& & -3\sig(x)\del(x-y)\sig(y)\dzmu\del(y-z)\gmu(z)\del(z-x)+ 
\nonumber \\ 
& & +\sig(x)\del(x-y)\sig(y)\del(y-z)\sig(z)\del(z-x)\gK + \dots 
\label{prop} 
\glausf
Integration by parts was used in Eq.~(\ref{prop}) to shift the 
derivatives from the
matrices containing the fields to the propagators.
In terms of the functions $A$ and $G_i$ 
defined in App.~\ref{app:loops}, the strong one-loop functional  
to $O(\phi^6)$ can be written 
\glanf 
&& \hspace*{-2cm}Z_4^{L=1}[\bar{U}, a, v, s, p] 
=\int \dx\,\gk \frac{1}{2}\,\sum_P \,A(M^2_P)\sig_{PP}(x)\nonumber \\ 
& &\hspace*{-2cm}\raum+\frac{1}{4}\,\sum_{P,Q}\,(A(M^2_P)+A(M^2_Q))  
{\grhou} \mbox{}_{PQ}(x)\grho_{QP}(x) \gK 
\nonumber \\ 
& & \hspace*{-2cm}+\int \dx\,\dy 
\frac{d^4p}{(2\pi)^4}\,\mbox{e}^{-i  
p (x-y)}\,\sum_{P,Q,R}\, 
\sgk G_{1\,PQ}^{\mu\nu}(p) \, 
\gmuu\mbox{}_{QP}(x)\gnuu\bk(y)  
\nonumber \\ 
& & \hspace*{-2cm}\raum+ G_{2\,PQ}^\mu(p) \,\gmuu\A(x)\sig\bk(y) 
\nonumber \\ 
& & \hspace*{-2cm}\raum 
+G_{3\,PQ}(p) 
\,\gk\sig\A(x)\sig\bk(y)+2\,\grhou\as(x)\grho\bs(x)\sig\cs(y)\gK\sgK 
\nonumber \\ 
& & \hspace*{-2cm}+\int \dx\,\dy\,\dz\frac{d^4 p_a}{(2\pi)^4}\frac{d^4 p_b}{(2\pi)^4} 
\,\mbox{e}^{i p_a (z-x)}\mbox{e}^{i p_b (y-z)}  
\nonumber \\ 
& & \hspace*{-2cm} \: \times\,\sum_{P,Q,R}\,\sgk G_{4\,PQR} ^{\mu\nu\lambda}(p_a,p_b)\, 
\gmuu\A(x)\gnuu\B(y)\glambdau\C(z) 
\nonumber \\ 
& & \hspace*{-2cm}\raum +G_{5\,PQR}^{\mu\nu}(p_a,p_b) \,\gmuu\A(x)\gnuu\B(y)\sig\C(z) 
\nonumber \\ 
& & \hspace*{-2cm} \raum +G_{6\,PQR}^{\mu}(p_a,p_b) 
\,\sig\A(x)\sig\B(y)\gmuu\C(z) 
\nonumber \\ 
& & \hspace*{-2cm}\raum  +G_{7\,PQR} (p_a,p_b)\sig\A(x)\sig\B(y)\sig\C(z)\sgK 
\label{eq:Zdiv} 
~.\glausf 
The divergent parts of the one-loop functional (\ref{eq:Zdiv}) are 
contained in the functions $A,B(0)$ defined in 
Eqs.~(\ref{eq:A},\ref{eq:B0}). Renormalization amounts to absorbing 
those divergences in the low-energy constants $L_i$ of the  
next-to-leading Lagrangian (\ref{eq:Lag4S}), rendering those constants  
scale dependent at the same time. In other words, to obtain the 
renormalized, scale-independent generating 
functional $Z_4$ in Eq.~(\ref{eq:Z4}), the divergences $\Lambda(\mu)$  
defined in (\ref{eq:Lambda}) are to be dropped in $Z_4^{L=1}$, with 
the $L_i$ being replaced by the renormalized constants $L_i^r(\mu)$ in  
the local functional $Z_4^{\rm tree}$. The analogous procedure will 
be understood for the nonleptonic weak part of the generating 
functional of $O(p^4)$ discussed in the following section. 
 
\begin{figure} 
\centerline{\begin{picture}(100,100) 
\put(50,50){\circle{40}} 
\put(70,50){\circle*{3}} 
\put(30,50){\circle*{3}} 
\put(70,50){\line(1,0){20}} 
\put(70,50){\vector(1,0){14}} 
\put(76.5,42){\makebox(10,10)[b]{\scriptsize $p$}} 
\put(30,50){\line(-1,0){20}} 
\put(8,50){\vector(1,0){14}} 
\put(14,42){\makebox(10,10)[b]{\scriptsize $p$}} 
\end{picture} 
\begin{picture}(100,100) 
\put(50,50){\circle{40}} 
\put(35.86,64.14){\circle*{3}} 
\put(35.86,35.86){\circle*{3}} 
\put(70,50){\circle*{3}} 
\put(35.86,64.14){\line(-1,1){14.14}} 
\put(35.86,35.86){\line(-1,-1){14.14}} 
\put(70,50){\line(1,0){20}} 
\put(21.72,78.28){\vector(1,-1){10}} 
\put(28.78,71.21){\makebox(8,8)[t]{\scriptsize $p_a$}} 
\put(35.86,35.86){\vector(-1,-1){10}} 
\put(28.78,18.78){\makebox(8,8)[r]{\scriptsize $p_b$}} 
\put(70,50){\vector(1,0){14}} 
\put(73,40){\makebox(25,8)[l]{\scriptsize $p_a\! -\!p_b$}} 
\end{picture}} 
\caption{Definition of external momenta.} 
\label{fig:momenta} 
\end{figure}
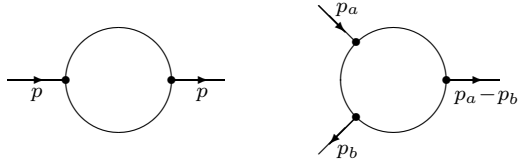 
The momentum flow shown in Fig.~\ref{fig:momenta} was chosen to 
fulfill the following criteria: 
\begin{itemize} 
\item The generating functional should contain as few terms as possible. 
\item The numerator of the three-point functions, i.e. 
$(k^2-M^2)$$((k-p_a)^2-M^2)$$((k-p_b)^2-M^2)$, should be symmetric under 
the interchange of $p_a$ and $p_b$. 
\end{itemize}

\section{Nonleptonic weak interactions} 
\label{sec:weak} 
\renewcommand{\theequation}{\arabic{section}.\arabic{equation}} 
\setcounter{equation}{0} 
To describe $|\Delta S|=1$ nonleptonic weak processes, one has to 
insert the nonleptonic weak Hamiltonian in the appropriate Green 
functions. At the hadronic level, this amounts to including exactly one 
vertex from the effective nonleptonic weak Lagrangian in the various 
parts of the generating functional.  
 
At leading order, the nonleptonic weak Lagrangian of $O(G_F \, p^2)$  
describing $|\Delta S|=1$ processes is given by \cite{Cronin:1967jq} 
\glan 
\cL_2^{W} =F^4\left[ G_8 \langle \lambda L_\mu L^\mu \rangle + 
G_{27}\left(L_{\mu 23} L^\mu_{11} + \frac{2}{3} L_{\mu 21} L^\mu_{13}\right) 
\right]+{\rm h.c.}\, , \label{eq:G827} 
\glaus 
$$ 
\lambda = (\lambda_6 - i \lambda_7)/2 ~, \raum 
L_\mu = i\, U^\dg D_\mu U~, 
$$ 
where $L_{\mu ij}$ denotes the $ij$-component of the matrix $L_\mu$. 
The weak mass term proportional to $\langle \lambda 
(U^\dg \chi + \chi^\dg U) \rangle$ can be transformed away by a 
suitable field redefinition \cite{Bernard:1985wf,Kambor:1989tz}.  
The octet and 27-plet coupling constants 
$G_8$ and $G_{27}$ can be extracted from $K \ra \pi \pi$ decay rates 
\cite{Cirigliano:2003gt}. We work consistently to first order in $G_8$ 
and $G_{27}$. 
 
To $O(p^2)$, with terms linear in $G_8, G_{27}$ included, the  
generating functional is still given by the 
classical action (\ref{eq:class}) with 
\begin{equation}  
\cL_2 = \cL_2^S + \cL_2^W ~. 
\end{equation}  
The classical solution $\bar{U}$ now satisfies a modified EOM where  
the weak part due to $\cL_2^W$ is added to (\ref{eq:EOM}). 
 
At $O(p^4)$, the structure of the generating functional is again given 
by Eq.~(\ref{eq:Z4}). The tree-level part $Z_4^{\rm tree}$ now 
contains the action for the nonleptonic weak Lagrangian of $O(G_F p^4)$ 
reproduced in Eqs.~(\ref{eq:L4W8},\ref{eq:L4W27}). The chiral anomaly  
affects weak 
amplitudes both through the weak part in the classical solution 
$\bar{U}$ in $Z_{\rm WZW}$ and intrinsically
through terms proportional to the $\epsilon$-tensor in the weak 
Lagrangian $\cL_4^W$ in (\ref{eq:L4W}) \cite{Ecker:1991bf}.  
 
It remains to determine the weak part of the one-loop functional 
$Z_4^{L=1}$. The operator $\delta$ defined in Eq.~(\ref{eq:ZTr}) 
is now of the form 
\begin{equation}  
\delta = \delta_S + \delta_W ~. 
\end{equation}  
The weak perturbation $\delta_W$ is given to first order in $G_F$ by 
the following expression \cite{Kambor:1989tz}: 
\glan \delta_W=\{\hat{\Gamma}_\mu^W,\partial^\mu \} 
+\{\hat{\Gamma}_\rho^W,\grho\}+\hat{\sigma}^W\, , \raum 
\hat{\Gamma}_\mu^W=-\frac{1}{2}\,N_\mu^- 
+\frac{1}{2}[T-T^\dag,\hat{\Gamma}_\mu]  \, , 
\glaus 
$$   
\hat{\sigma}^W=\hat{\omega}+ 
\frac{1}{2}[d^\mu,N^+_\mu]-\frac{1}{2}[d^\mu,[d_\mu,\alpha]]- 
\frac{1}{2}\{\bar{\alpha},\hat{\sigma}\}+T\bar{\sigma}+\bar{\sigma}T ^\dag
\, ,  \raum  
$$ 
$$ 
T_{ab}=\frac{\hat{\sigma}^0_{ac}\, \alpha^0_{cb}}{M^2_b-M^2_a} \, , \raum 
d_\mu A=\partial_\mu A+[\gmu, A] \, , \raum \bar{\alpha}=\alpha-\alpha^0 \, . 
$$ 
The matrices $\hat{\sigma}^{0}$ and $\alpha^0$ are obtained from 
$\hat{\sigma}$ and $\alpha$ by switching off the external fields. 
In the weak perturbation there are no terms without external fields  
because a diagonalizing transformation similar to the one in  
Ref.~\cite{Ecker:1987hd} has been performed.  
The components of the antisymmetric matrix $N^-_{\mu}$ are  
\glanf N^-_{\mu a b} &=& -\frac{1}{4}\kappa \kk\{K_{32},y_\mu\} 
[\lambda_a,\lambda_b]\KK 
-\frac{1}{2}\kappa \kk K_{32} \left(\lambda_a\, y_\mu\, \lambda_b- 
\lambda_b\, y_\mu\, \lambda_a \right) \KK  
\label{eq:Nminus} \\ 
&& -\zeta\kk [\lambda_a,\lambda_b ]K\KK\kk K y_\mu\KK 
+\frac{1}{2}\zeta\kk \lambda_a [K, y_\mu]\KK \kk K \lambda_b \KK 
-\frac{1}{2}\zeta\kk \lambda_b [K, y_\mu]\KK \kk K \lambda_a \KK \,  
+ {\rm h.c.} \nn 
\glausf 
The coefficient $\kappa$ stands for  
\glan \kappa=4\,G_8 \, F^2 
\glaus 
and the matrix K is defined by matrix elements 
\glan K_{ij}=\bar{u} \lambda_{ij} \bar{u}^\dag  \quad 
{\rm with} \quad (\lambda_{ij})_{kl}=\delta_{ik} \delta_{jl} ~. 
\glaus 
The operators $\zeta \kk K {\cal O}\KK \kk K {\cal P}\KK $ are due to the 
27-plet contribution and are defined as 
\glanf  
&& \zeta \kk K {\cal O}\KK \kk K {\cal P}\KK =  
4 \, G_{27}  F^2 \,t_{ij,kl} \kk K_{ij}  
{\cal O}\KK \kk K_{kl} {\cal P}\KK ~,\label{eq:qKK} 
\glausf 
with coefficients 
\glanf && t_{23,11}=t_{11,23}=\frac{1}{2} \, , \raum 
t_{21,13}=t_{13,21}=\frac{1}{3}\, , \nl  
&& t_{ij,kl}=0 \raum\mbox{otherwise} ~. \label{eq:tijkl} 
\glausf 
The expressions for the remaining symmetric matrices are 
\glanf \vspace{-0.5cm} 
N^+_{\mu a b} &=& \frac{1}{4}\kappa \kk [K_{32},y_\mu ] 
\{\lambda_a,\lambda_b\}\KK \nl 
&& +\frac{1}{2}\zeta\kk \lambda_a [K, y_\mu]\KK \kk K \lambda_b \KK  
 +\frac{1}{2}\zeta\kk \lambda_b [K, y_\mu]\KK \kk K \lambda_a \KK  
+ {\rm h.c.}\, ,  
\glausf 
\glanf \hat{\omega}_{a b} &=& -\frac{1}{4}\kappa\kk\{\lambda_a, 
\lambda_b\}\{y^2,K_{32}\}\KK 
+\frac{1}{4}\kappa\kk K_{32} \left( \lambda_a y^2 \lambda_b  
+ \lambda_b y^2 \lambda_a \right) \KK 
\nonumber \\ 
&& +\frac{1}{4}\kappa\kk \{K_{32},y^\rho\}(\lambda_a\, y_\rho \lambda_b 
+\lambda_b\, y_\rho \lambda_a )\KK  
-\frac{1}{4}\kappa\kk y^\rho K_{32} y_\rho \{\lambda_a , \lambda_b\}\KK 
\nonumber \\ 
&& -\frac{1}{2}\zeta \kk [K, y^\rho]\lambda_a\KK \kk [K, y_\rho]\lambda_b\KK 
-\frac{1}{2}\zeta\kk K y^\rho\KK \kk \{ K, y_\rho\} \{\lambda_a , \lambda_b\}\KK 
\nonumber \\ 
&& +\zeta\kk K \left( \lambda_a y^\rho \lambda_b + \lambda_b y^\rho \lambda_a  
\right) \KK \kk K y_\rho \KK\,+ {\rm h.c.} , 
\glausf 
\glanf \alpha_{a b}=\frac{1}{4}\kappa\kk\{\lambda_a , \lambda_b\} 
K_{32}\KK +\frac{1}{2}\zeta \kk \lambda_a K \KK \kk \lambda_b K\KK \,  
+ {\rm h.c.} 
\glausf 
To get the weak functional one must replace in \glei{eq:Zdiv} 
$\gmuu$ by $\hat{\Gamma}_\mu^W$ and $\sig$ by  
$\hat{\sigma}^W$ once in each term and insert the 
appropriate binomial factors.  
 
\section{Mathematica$^{\textrm{\scriptsize \copyright}}$  
program Ampcalculator} 
\label{sec:Math} 
\renewcommand{\theequation}{\arabic{section}.\arabic{equation}} 
\setcounter{equation}{0} 
 
The program Ampcalculator \cite{ampcalc} evaluates the complete 
amplitudes for strong and nonleptonic weak mesonic processes to  
$O(p^4)$ and to $O(\phi^6)$ (counting an external photon as 
$O(\phi^2)$), with the following exceptions. 
\begin{itemize}  
\item In the Wess-Zumino-Witten functional \cite{Wess:1971yu} 
  incorporating the chiral anomaly, only terms with explicit spin-1 
  fields are processed. In other words, we omit for practical 
  reasons\footnote{The associated vertices contain at least five meson  
  fields.} the part that cannot be written as the integral of a finite 
  polynomial in $U$, $\partial_\mu U$ in four dimensions but can be 
  written as a five-dimensional integral, with Minkowski space as 
  boundary of the five-dimensional manifold. 
\item For radiative semileptonic decays (e.g., for $K^+ \to l^+ \nu_l 
  \gamma$), the Bremsstrahlung amplitude(s) for the photon(s) coupling  
  to the charged lepton must be added by hand. For instance, for a 
  single photon in the final state the program only calculates 
  hadronic matrix elements of the type 
\begin{equation}  
M_{\mu\nu} = i \displaystyle\int d^4 x \,e^{\displaystyle i q.x} \langle 
{\rm hadrons}|T \, V_\mu^{\rm elm}(x) \left(V_\nu(0) - A_\nu(0)  
\right)^{|\Delta S|=1}| {\rm M(eson)} \rangle  
\end{equation} 
and then multiplies the result with the appropriate factors to get the 
(partial) amplitude for the decay $M \to {\rm hadrons}~+l +\nu_l + 
\gamma$. 
\end{itemize}

By default, the program reduces the one-loop amplitudes down to the  
basic loop functions $A,B,C$ defined in App.~\ref{app:loops}. 
In certain cases, the result may be more transparent and even more 
suitable for numerical treatment without performing the 
recursion relations for the various loop integrals collected in 
App.~\ref{app:loops}. The user has the option to skip the recursion 
relations by setting the variable norecurs=1. 
 
For running Ampcalculator, a list of external mesons, photons and 
$W^\pm$ must be specified together with their momenta in an input file 
(cf. App.~\ref{app:input}). All particles are assumed to be incoming.  
The program 
checks for charge conservation and exits if the total charge of the 
listed particles is non-zero. For convenience and better readability 
of the output, the user can also provide two lists with scalar 
products of momenta and polarization vectors (also used 
for virtual photons for simplicity).  
The first list contains the scalar products themselves and the 
second one the corresponding scalar variables. In this way, momentum 
conservation can be implemented. The lists can also be used to redefine 
momenta (see Sec.~\ref{subsec:Kl3} for an example). 
For semileptonic transitions, the 
program calculates the $V-A$ hadronic amplitude and then multiplies 
with the appropriate factors to get the full amplitude (except for 
lepton Bremsstrahlung) including the 
lepton matrix elements $l_\mu$ or $\hat{l}_\mu$, where 
\begin{eqnarray} 
l_\mu &=& \bar{u}(p_{\nu_l}) \gamma_\mu (1-\gamma_5) v(p_{l^+})~, \nl 
\hat{l}_\mu &=& \bar{u}(p_{l^-}) \gamma_\mu (1-\gamma_5) 
v(p_{\bar{\nu_l}}) ~. 
\label{eq:leptons} 
\end{eqnarray}  
 
The masses of the pseudoscalar mesons 
(always in the isospin limit) need not be specified explicitly 
for the squares of meson momenta. However, the 
user should specify (for his/her own convenience) if the external 
photon(s) is (are) real: for an on-shell photon with momentum $k$ and  
polarization vector $\ve[k]$, the lists should contain the 
declarations for $k \cdot k=k \cdot \ve[k] = 0$ 
(cf., e.g., Sec.~\ref{subsec:Kpgg}). In the amplitudes of $O(p^4)$,  
by default $M_\eta^2$ is replaced using the 
Gell-Mann--Okubo relation  
\begin{equation} 
M_\eta^2 = \displaystyle\frac{4}{3}\left(4 M_K^2 - M_\pi^2 \right)~,   
\end{equation}  
except as argument of a loop function where for better readability  
$M_\eta^2$ is always kept. The user has the option to keep  
$M_\eta^2$ everywhere. 
The complex conjugate of a quantity $G$ (e.g., 
$G_8$, $N_i$, $V_{us}$, \dots) is denoted generically as $\hat{G}$.   
 
Running Ampcalculator may be quite time consuming, especially for  
nonleptonic weak processes,  
the time increasing strongly with the number of  
external states. However, the big advantage is that it is computer  
time rather than the physicist's time that is being 
consumed. Especially for exploratory purposes, the program provides 
options for calculating the tree amplitudes (at lowest order or to 
$O(p^4)$) and the loop amplitudes separately. Details can be found in 
App.~\ref{app:input} where the input file for Ampcalculator is 
listed. The input file ampcalculator.nb and the subroutine 
ampcalculatorsub.nb are available for general use \cite{ampcalc}. The 
results are written into a separate output file.

To perform CHPT calculations, another Mathematica$^{\textrm{\scriptsize
\copyright}}$ package, called PHI \cite{PHI}, exists. That package makes
use of the more general package FeynCalc \cite{Kublbeck:1992mt}. In
contrast to Ampcalculator, FeynCalc/PHI is not a fully automatic
program. In order to use
FeynCalc/PHI, detailed knowledge of Mathematica$^{\textrm{\scriptsize
\copyright}}$ and of the packages FeynCalc and PHI is needed.
Another difference between FeynCalc/PHI and Ampcalculator is the way the
amplitudes are calculated. FeynCalc/PHI generates and uses Feynman rules
whereas Ampcalculator performs a functional differentiation of the
generating functional. Both packages are completely
independent allowing for an excellent check of the results.
 
In the remainder of this section we present input and output for three  
examples. Although the examples are nontrivial 
from the point-of-view of computer time needed the amplitudes are 
relatively simple and can easily be compared with  
results available in the literature. The indicated execution times  
were obtained running Mathematica$^{\textrm{\scriptsize \copyright}}$5 
on a PC with 512 MB RAM at 1.6 GHz.

In addition to the three examples shown below, we have checked the
existing results for the processes $\gamma \gamma \to \pi^+ \pi^-$ 
\cite{Bijnens:1987dc}, $e^+ e^- \to 4 \pi$ 
\cite{Unterdorfer:2002zg, Ecker:2002cw}, $K^+ \pi^+ \to K^+ \pi^+$ 
\cite{Bernard:1990kw} and $K^0 \to 3 \pi^0$ 
\cite{Kambor:1991ah, Bijnens:2002vr}. 
 
\subsection{The semileptonic decay $K^+ \to \pi^0 l^+ \nu_l 
  ~(K^+_{l3})$} \label{subsec:Kl3} 
The program Ampcalculator evaluates the hadronic $V - A$ amplitude 
and then multiplies it with the appropriate factor to get the full  
decay amplitude. With the external momenta defined through 
$$ 
K^+ (p) \to \pi^0 (p^\prime) l^+ (p_l) \nu_l (p_\nu) 
$$ 
with 
\begin{eqnarray}  
q=p_l + p_\nu~= p - p^\prime, \quad  P=p + p^\prime~, 
\end{eqnarray}  
a possible input for the lists ExternalParticles, replist1 and 
replist2 is given below. Of course, the choice of momenta and kinematic  
variables is up to the user. Remember that the program interprets all 
particles as incoming. \\ \\ 
{\fontfamily{phv}\selectfont ExternalParticles 
= \{K$_+$[p], W$_-$[Q], $\pi_0$[r]\};} \\ \\ 
{\fontfamily{phv}\selectfont replist1 
= \{sp[Q, Q], sp[r, Q], sp[p, Q], sp[p, r], p$^\mu$, r$^\mu$,  
Q$^\mu$\};} \\ \\ 
{\fontfamily{phv}\selectfont replist2 
= \{t, (M$_K^2$$-$M$_\pi^2-$t)/2, ($-$M$_K^2$$+$M$_\pi^2-$t)/2, 
($-$M$_K^2$$-$M$_\pi^2+$t)/2, (P$^{\mu}+$q$^\mu$)/2,  
($-$P$^{\mu}+$q$^\mu$)/2, $-$q$^\mu$\};} \\ \\ 
The complete result of $O(p^4)$ is printed out in a separate output 
file (execution time: 0.30 hours). \\ \\ 
{\fontfamily{phv}\selectfont Complete amplitude to O(p\^{}4)} 
\glanf && 
\mathtt{{G_F}\,{l_{\mu }}\,{{{\hat{V}}}_ 
    {{us}}}\, 
  \sgk   
   \frac{P^{\mu }}{2} + \frac{ 2\,q^{\mu }\,L_5^r\, 
       \left( {{M_K}}^2 - {{M_{\pi }}}^2 \right) }{ 
         {{F_{\pi }}}^2} +  
    \frac{L_9^r\,\left( P^{\mu }\,t +  
         q^{\mu }\,\left( -{{M_K}}^2 +  
            {{M_{\pi }}}^2 \right)  \right) }{{{F_ 
          {\pi }}}^2}} \nl && +  
   \mathtt{ \frac{\left( t - 5\,{{M_K}}^2 - {{M_{\pi }}}^2 
         \right) \,\left( P^{\mu }\,t +  
         q^{\mu }\,\left( -{{M_K}}^2 +  
            {{M_{\pi }}}^2 \right)  \right) }{192\, 
       {\pi }^2\,t\,{{F_{\pi }}}^2} } \nl && -  
    \mathtt{\frac{\left( -2\,q^{\mu }\, 
          \left( {{M_K}}^2 - {{M_{\pi }}}^2 \right) \, 
          \left( t - 4\,{{M_K}}^2 +  
            2\,{{M_{\pi }}}^2 \right)  +  
         P^{\mu }\,t\,\left( t - 8\,{{M_K}}^2 +  
            4\,{{M_{\pi }}}^2 \right)  \right) \, 
       {\bar{A}}[{{M_K}}^2]}{8\,t\,{{F_{\pi }}}^2\,  
       \left( {{M_K}}^2 - {{M_{\pi }}}^2 \right) } }\nl &&-  
     \mathtt{\frac{\left( P^{\mu }\,t\, 
          \left( t - 4\,{{M_K}}^2 \right)  +  
         4\,q^{\mu }\,\left( t + {{M_K}}^2 \right) \, 
          \left( {{M_K}}^2 - {{M_{\pi }}}^2 \right)  
         \right) \,{\bar{A}}[{{M_{\pi }}}^2]}{16\,t\, 
       {{F_{\pi }}}^2\, 
       \left( {{M_K}}^2 - {{M_{\pi }}}^2 \right) } }\nl && +  
    \mathtt{\frac{3\,\left( P^{\mu }\,t\, 
          \left( t - 4\,{{M_K}}^2 \right)  +  
         4\,q^{\mu }\,{{M_K}}^2\, 
          \left( {{M_K}}^2 - {{M_{\pi }}}^2 \right)  
         \right) \,{\bar{A}}[{{M_{\eta }}}^2]}{16\,t\, 
       {{F_{\pi }}}^2\, 
       \left( {{M_K}}^2 - {{M_{\pi }}}^2 \right) } }\nl &&+  
    \mathtt{ \sgk P^{\mu }\,t\, 
          \left( {{M_K}}^4 +  
            {\left( 3\,t + {{M_{\pi }}}^2 \right) }^2 -  
            2\,{{M_K}}^2\, 
             \left( 21\,t + {{M_{\pi }}}^2 \right)  
            \right)  }\nl && -  
         \mathtt{ 4\,q^{\mu }\,\left( {{M_K}}^2 -  
            {{M_{\pi }}}^2 \right) \, 
          \left( {{M_K}}^4 + 3\,t\,{{M_{\pi }}}^2 +  
            {{M_{\pi }}}^4 -  
            {{M_K}}^2\, 
             \left( 9\,t + 2\,{{M_{\pi }}}^2 \right)  
            \right)  \sgK \,\, 
       \frac{\bar{B}[t,{{M_K}}^2,{{M_{\eta }}}^2]}{144\,t^2\, 
       {{F_{\pi }}}^2} }\nl && +  
    \mathtt{\sgk -4\,q^{\mu }\, 
          \left( {{M_K}}^2 - {{M_{\pi }}}^2 \right) \, 
          \left( -t^2 + {{M_K}}^4 -  
            2\,{{M_K}}^2\,{{M_{\pi }}}^2 +  
            {{M_{\pi }}}^4 \right) }\nl && +  
         \mathtt{ P^{\mu }\,t\,\left( {{M_K}}^4 +  
            {\left( t - {{M_{\pi }}}^2 \right) }^2 -  
            2\,{{M_K}}^2\, 
             \left( t + {{M_{\pi }}}^2 \right)  \right)  
         \sgK \,\frac{{\bar{B}}[t,{{M_{\pi }}}^2,{{M_K}}^2]} 
       {16\,t^2\,{{F_{\pi }}}^2} } \sgK  \nonumber 
\glausf 
The result agrees with Ref.~\cite{Gasser:1984ux}. 
 
\subsection{The nonleptonic decay $K^- \to \pi^- \pi^0$} 
 
In this decay, the complex conjugate coupling constants 
$\hat{G}_{27}$ and $\hat{R}^r_i$ appear in the amplitude. The 
assignment of momenta and scalar products is straightforward. \\ \\ 
{\fontfamily{phv}\selectfont ExternalParticles 
= \{K$_-$[p$_1$], $\pi_+$[q$_1$], $\pi_0$[q$_2$]\};} \\ \\ 
{\fontfamily{phv}\selectfont replist1 
= \{sp[p$_1$, q$_1$], sp[p$_1$, q$_2$], sp[q$_1$, q$_2$]\};} \\ \\ 
{\fontfamily{phv}\selectfont replist2 
= \{-$\frac{1}{2}\,$M$_K^2$,\,-$\frac{1}{2}\,$M$_K^2$,\, 
-M$_\pi^2$+$\,\frac{1}{2}\,$M$_K^2$\};} \\ \\ 
{\fontfamily{phv}\selectfont Complete amplitude to O(p\^{}4)} 
\glanf 
&& \mathtt{i \,\hat{G}_{27}\sgk\frac{5 }{3}\,{F_{\pi }}\, 
   \left( {{M_K}}^2 - {{M_{\pi }}}^2 \right)  -  
  \frac{80\, L_4^r\, 
     \left( 2\,{{M_K}}^4 - {{M_K}}^2\,{{M_{\pi }}}^2 -  
       {{M_{\pi }}}^4 \right) }{{3\,F_{\pi }}} 
   }\nl && -  
  \mathtt{\frac{20\, L^r_5\, 
     \left( {{M_K}}^4 + 2\,{{M_K}}^2\,{{M_{\pi }}}^2 -  
       3\,{{M_{\pi }}}^4 \right) }{{3\,F_{\pi }}}  +  
  \frac{5\,  
     \left( {{M_K}}^4 - 3\,{{M_K}}^2\,{{M_{\pi }}}^2 +  
       2\,{{M_{\pi }}}^4 \right) }{{96\,\,{\pi }^2\,F_{\pi }}} }\nl && 
   + \mathtt{\frac{5\,  
     \left( {{M_K}}^4 + 3\,{{M_K}}^2\,{{M_{\pi }}}^2 -  
       4\,{{M_{\pi }}}^4 \right) \,\hat{R}_8^r}{{3\,F_{\pi }}}  
   + \frac{5\,  
     \left( {{M_K}}^4 - {{M_K}}^2\,{{M_{\pi }}}^2 
       \right) \,\hat{R}_9^r}{{3\,\,F_{\pi }}} }\nl &&+ %
  \mathtt{\frac{10\,  
     \left( {{M_K}}^2 - {{M_{\pi }}}^2 \right) \, 
     \left( 2\,{{M_K}}^2 + {{M_{\pi }}}^2 \right) \, 
     \hat{R}_{10}^r}{{3\,\,F_{\pi }}} +  
  \frac{20\, {{M_{\pi }}}^2\, 
     \left( {{M_K}}^2 - {{M_{\pi }}}^2 \right) \, 
     \hat{R}_{12}^r}{{3\,\,F_{\pi }}} }\nl &&+  
  \mathtt{\frac{5\,  
     \left( 3\,{{M_K}}^2 + {{M_{\pi }}}^2 \right) \, 
     {\bar{A}}[{{M_K}}^2]}{{12\,\,F_{\pi }}} -  %
  \frac{5\,  
     \left( 4\,{{M_K}}^4 -  
       22\,{{M_K}}^2\,{{M_{\pi }}}^2 +  
       29\,{{M_{\pi }}}^4 \right) \, 
     {\bar{A}}[{{M_{\pi }}}^2]}{24\,\,{F_{\pi }}\, 
     {{M_{\pi }}}^2} }\nl &&+  
  \mathtt{\frac{5\, {{M_{\pi }}}^2\, 
     {\bar{A}}[{{M_{\eta }}}^2]}{{8\,\,F_{\pi }}} -  
  \frac{5\,  
     \left( {{M_K}}^4 - 3\,{{M_K}}^2\,{{M_{\pi }}}^2 +  
       2\,{{M_{\pi }}}^4 \right) \, 
     {\bar{B}}[{{M_K}}^2,{{M_{\pi }}}^2,{{M_{\pi }}}^2]} 
     {6\,\,{F_{\pi }}} }\nl &&+ \mathtt{\frac{5\,  
     {{M_K}}^4\, 
     \left( -{{M_K}}^2 + {{M_{\pi }}}^2 \right) \, 
     {\bar{B}}[{{M_{\pi }}}^2,{{M_K}}^2,{{M_{\eta }}}^2]} 
     {72\,\,{F_{\pi }}\,{{M_{\pi }}}^2} }\nl &&-  
  \mathtt{\frac{5\, {{M_K}}^2\, 
     \left( 5\,{{M_K}}^4 -  
       13\,{{M_K}}^2\,{{M_{\pi }}}^2 + 8\,{{M_{\pi }}}^4 
       \right) \,{\bar{B}}[{{M_{\pi }}}^2,{{M_{\pi }}}^2, 
      {{M_K}}^2]}{24\,\,{F_{\pi }}\,{{M_{\pi }}}^2}\sgK \nonumber}
\glausf 
 
With the appropriate renaming of coupling constants, 
the result (execution time: 1.48 hours) agrees with previous  
calculations in Ref.~\cite{Kambor:1991ah, Bijnens:1998mb}.

\subsection{The nonleptonic decay $K^0 \to \pi^0 \gamma \gamma$} 
\label{subsec:Kpgg} 
Finally, we present the amplitude for a decay where three-propagator 
loops contribute: 
$$ 
K^0 (p) \to \pi^0 (p^\prime) \gamma (q_1) \gamma (q_2) ~.  
$$ 
In the limit of CP conservation, the even-intrinsic-parity amplitude  
determines the amplitude for $K_L  \to \pi^0 \gamma \gamma$  
whereas the part with the $\epsilon$-tensor yields the  
corresponding $K_S$  
amplitude. The following assignments (all particles incoming) 
correspond to $p_1= - p^\prime$, $k_1= - q_1$, $k_2 = - q_2$. \\ \\ 
{\fontfamily{phv}\selectfont ExternalParticles 
= \{K$_0$[p], $\pi_0$[$p_1$], $\gamma$[$k_1$],  
$\gamma$[$k_2$]\};} \\ \\ 
{\fontfamily{phv}\selectfont replist1 
= \{sp[p,p$_1$], sp[k$_1$,k$_2$], sp[k$_1$,$\epsilon$[k$_1$]],  
  sp[k$_1$,$\epsilon$[k$_2$]], sp[k$_2$,$\epsilon$[k$_1$]],  
  sp[k$_2$,$\epsilon$[k$_2$]], sp[$\epsilon$[k$_1$],$\epsilon$[k$_2$]],  
  sp[p,$\epsilon$[k$_1$]], sp[p,$\epsilon$[k$_2$]],  
  sp[p$_1$,$\epsilon$[k$_1$]], sp[p$_1$,$\epsilon$[k$_2$]],  
  sp[k$_1$,k$_1$], sp[k$_2$,k$_2$]\};} \\ \\ 
{\fontfamily{phv}\selectfont replist2 
= \{$\frac{1}{2}$ ($s$-M$_K^2$-M$_\pi^2$), $\frac{s}{2}$, 
  0, -q$_1$$\cdot$$\,\epsilon _2$, -q$_2$$\cdot$$\,\epsilon_1$, 0,  
  $\epsilon_1$$\cdot$$\,\epsilon_2$, p$\cdot$$\,\epsilon_1$, p$\cdot$$\,\epsilon_2$,  
  -p$\cdot$$\,\epsilon_1$+q$_2$$\cdot$$\,\epsilon_1$,  
  -p$\cdot$$\,\epsilon_2$+q$_1$$\cdot$$\,\epsilon_2$, 
  0, 0\};} \\ \\ 
{\fontfamily{phv}\selectfont Complete amplitude to O(p\^{}4)} 
\glanf 
&& \mathtt{\frac{1}{2}e^2\sgk G_8\,\sgk\frac{\left( 2\,{q_1}\cdot {{\epsilon }_2}\, 
        {q_2}\cdot {{\epsilon }_1} -  
       s\,{{\epsilon }_1}\cdot {{\epsilon }_2} \right) 
      {{M_K}}^2 }{4\,{\sqrt{2}}\,{\pi }^2\,s} }\nl &&+  
  \mathtt{\frac{4 
      }{s} {\sqrt{2}}\, 
     {C}[0,0,-\frac{s}{2},{{M_{\pi }}}^2, 
      {{M_{\pi }}}^2,{{M_{\pi }}}^2]\, 
     \left( 2\,{q_1}\cdot {{\epsilon }_2}\, 
        {q_2}\cdot {{\epsilon }_1} -  
       \Mfunction{s}\,{{\epsilon }_1}\cdot  
         {{\epsilon }_2} \right) \,{{{{M}}_{\pi }}}^2\, \left( s - {{M_{\pi }}}^2 \right)}\nl && 
\nl &&+  
  \mathtt{\frac{4 
       }{s\,\left( {{M_K}}^2 - {{M_{\pi }}}^2 \right) 
       } {\sqrt{2}}\, 
     \Mfunction{C}[0,0,-\frac{s}{2},{{M_K}}^2,{{M_K}}^2, 
      {{M_K}}^2]\,\left( -2\, 
        {{\Mfunction{q}}_1}\cdot {{\epsilon }_2}\, 
        {{\Mfunction{q}}_2}\cdot {{\epsilon }_1} +  
       \Mfunction{s}\,{{\epsilon }_1}\cdot  
         {{\epsilon }_2} \right) \,{{{\Mfunction{M}}_K}}^2} \nl && 
      \mathtt{ \times\left( s\,{{M_K}}^2 - {{M_K}}^4 -  
          s\,{{M_{\pi }}}^2 + {{M_{\pi }}}^4 \right)  } 
         \sgK\nl && 
   +\mathtt{G_{27}\sgk\frac{\left( 2\,{q_1}\cdot {{\epsilon }_2}\, 
        {q_2}\cdot {{\epsilon }_1} -  
       s\,{{\epsilon }_1}\cdot {{\epsilon }_2} \right) 
           \left( -13\,{{M_K}}^2 +  
          15\,\left( s - {{M_{\pi }}}^2 \right)  \right) 
            }{12\,{\sqrt{2}}\,{\pi }^2\,s} }\nl &&+  
  \mathtt{\frac{1 
      }{3\,s\, 
     \left( {{M_K}}^2 - {{M_{\pi }}}^2 \right) }  
     {2\sqrt{2}}\, 
     {C}[0,0,-\frac{s}{2},{{M_{\pi }}}^2, 
      {{M_{\pi }}}^2,{{M_{\pi }}}^2]\, 
     \left( 2\,{q_1}\cdot {{\epsilon }_2}\, 
        {q_2}\cdot {{\epsilon }_1} -  
       \Mfunction{s}\,{{\epsilon }_1}\cdot  
         {{\epsilon }_2} \right) \, 
     {{{{M}}_{\pi }}}^2\, }\nl && 
   \mathtt{ 
     \times 
        \left( -5\,{{M_K}}^4 +  
          24\,{{M_{\pi }}}^2\, 
           \left( -s + {{M_{\pi }}}^2 \right)  +  
          {{M_K}}^2\,\left( 9\,s + {{M_{\pi }}}^2 
             \right)  \right)  
    }\nl &&+  
  \mathtt{\frac{1 
       }{3\,s\,\left( {{M_K}}^2 - {{M_{\pi }}}^2 \right) 
       }  
       {2\sqrt{2}}\, 
     \Mfunction{C}[0,0,-\frac{s}{2},{{M_K}}^2,{{M_K}}^2, 
      {{M_K}}^2]\,\left( -2\, 
        {{\Mfunction{q}}_1}\cdot {{\epsilon }_2}\, 
        {{\Mfunction{q}}_2}\cdot {{\epsilon }_1} +  
       \Mfunction{s}\,{{\epsilon }_1}\cdot  
         {{\epsilon }_2} \right) \,{{{\Mfunction{M}}_K}}^2\, 
       }\nl && 
   \mathtt{  
     \times 
    \left( 21\,{{M_K}}^4 +  
          6\,{{M_{\pi }}}^2\, 
           \left( s - {{M_{\pi }}}^2 \right)  +  
          {{M_K}}^2\,\left( -21\,s +  
             5\,{{M_{\pi }}}^2 \right)  \right)} \sgK\sgK\nl &&     
       + \mathtt{ \frac{1}{2}\, i\,e^2 \sgk \frac{2{\sqrt{2}}\, 
     {\epsilon }^{\Mvariable{\xi \rho \sigma \tau }}\, 
     {G_8}\, 
     \left( s - {{M_K}}^2 \right) \, 
     \left( {{M_K}}^2 - {{M_{\pi }}}^2 \right) \, 
     \left( {p_{\xi }}\,{{{k_1}}_{\rho }} -  
       {{{k_1}}_{\xi }}\,{{{p_1}}_{\rho }} \right) \, 
     {{\left( \epsilon ({k_1}) \right) }_{\sigma }}\, 
     {{\left( \epsilon ({k_2}) \right) }_{\tau }}}{ 
     {\pi }^2\,\left( s - {{M_{\pi }}}^2 \right) \, 
     \left( 3\,s - 4\,{{M_K}}^2 + {{M_{\pi }}}^2 \right) 
       } } 
       \nl &&     
       - \mathtt{\frac{2{\sqrt{2}}\, 
     {\epsilon }^{\Mvariable{\xi \rho \sigma \tau }}\, 
     {G_{27}}\, 
     \left( s - {{M_K}}^2 \right) \, 
     \left( {{M_K}}^2 - {{M_{\pi }}}^2 \right) \, 
     \left( {p_{\xi }}\,{{{k_1}}_{\rho }} -  
       {{{k_1}}_{\xi }}\,{{{p_1}}_{\rho }} \right) \, 
     {{\left( \epsilon ({k_1}) \right) }_{\sigma }}\, 
     {{\left( \epsilon ({k_2}) \right) }_{\tau }}}{ 
     {\pi }^2\,\left( s - {{M_{\pi }}}^2 \right) \, 
     \left( 3\,s - 4\,{{M_K}}^2 + {{M_{\pi }}}^2 \right)  
       } \sgK} 
       \nonumber 
\glausf 
 
The amplitude (execution time: 1.68 hours) agrees with what is  
available in the literature  
\cite{Ecker:1987fm,Cappiello:1988yg}. The function $F$ defined in  
Ref.~\cite{Ecker:1987fm} is related to the three-propagator loop  
function as 
\begin{equation} 
F(s/M^2) = 1 + 32 \pi^2\,C(0,0,- s/2,M^2,M^2,M^2)~. 
\end{equation} 
In the anomaly contribution, $M_\eta^2$ was replaced via the  
Gell-Mann--Okubo relation. To our knowledge (and for good 
reasons), the 27-plet kaon loop amplitude and the anomalous 
contribution proportional to $G_{27}$ have not been published before.

\section{Conclusions} 
\label{sec:concs} 
\renewcommand{\theequation}{\arabic{section}.\arabic{equation}} 
\setcounter{equation}{0} 
The generating functional is a convenient quantity for a compact 
representation of Green functions and S-matrix elements. Especially for 
an effective field theory like chiral perturbation theory, with its 
many coupling constants and derivative couplings, the standard 
diagrammatic calculations can be quite cumbersome and they must be 
performed for each process separately. The great advantage of the 
generating functional is that the renormalized amplitudes can be  
obtained once and for all. 
 
Extending previous work, we have presented in this paper the 
generating functional of Green functions for chiral SU(3) for both  
strong and nonleptonic weak interactions in the meson sector to  
next-to-leading order in the low-energy expansion and for at most six  
external states, a photon counting as $O(\phi^2)$. Such processes 
require the inclusion of one-loop diagrams with up to three meson  
propagators and they comprise almost all strong and nonleptonic weak  
mesonic transitions of physical interest. 
 
The main purpose of this work is to reproduce and to check previous 
calculations in a simple and straightforward way but the functional 
presented here has also been used to derive new results 
\cite{Isidori:2004rb}. Although the representation is compact the 
actual extraction of a specific amplitude via three-dimensional 
flavour traces can be quite laborious, the toil increasing of course 
with the number of external states and especially for nonleptonic weak 
transitions. We have therefore provided as an essential part of this 
work the Mathematica$^{\textrm{\scriptsize \copyright}}$ program  
Ampcalculator for general use that performs the necessary 
manipulations upon input of the relevant external particles with their  
momenta. 
 
We have described the necessary input for the calculation of three 
semileptonic and nonleptonic $K$ decay amplitudes and presented the  
resulting output. As far as available, the results agree with previous 
calculations.  
 
The generating functional presented here can be extended in several
ways. In addition to radiative corrections for strong processes 
\cite{Schweizer:2002ft,schweizersu3}, photons and leptons can be
included as dynamical degrees of freedom for similar treatments of 
radiative corrections for semileptonic
\cite{Knecht:1999ag} and nonleptonic weak decays.

\vfill 
\vspace*{1cm}  
\section*{Acknowledgements} 
\noindent 
We are grateful to Julia Schweizer and Roland Rosenfelder for a critical reading of the 
manuscript. R.U. wishes to express his thanks to Giulia Pancheri and 
Gino Isidori for the hospitality during his stay in Frascati. 
 
\appendix  
 
\newcounter{zaehler} 
\renewcommand{\thesection}{\Alph{zaehler}} 
\renewcommand{\theequation}{\Alph{zaehler}.\arabic{equation}}

\setcounter{equation}{0} 
\addtocounter{zaehler}{1} 
\section{Input for the program Ampcalculator} 
\label{app:input} 
 
We list here the contents of the input file ampcalculator.nb. During  
execution of the input file, the subroutine ampcalculatorsub.nb is 
called where the actual computations are performed. 
The result is then printed out in a separate output file.

\texttt{ 
\begin{center}  
Program  Ampcalculator  
\end{center} 
The program generates the amplitude for any strong or nonleptonic weak  
process of pseudoscalar mesons, external photons and W bosons to  
next-to-leading order in the low-energy expansion (chiral SU(3)) with 
at most six external particles (a photon counts as two particles), at  
most one W (semileptonic decays) and with at most three propagators in  
the loop. The W is assumed to couple to a lepton pair indicated in the  
output by the leptonic matrix elements $\mathtt{l_\mu}$ 
and $\mathtt{\hat{l}_\mu}$, defined by 
\begin{eqnarray*} 
\mathtt{l_\mu} = \mathtt{\bar{u}(p_{\nu}) \gamma_\mu (1-\gamma_5)  
v(p_{l^+})~, } & \quad & 
\mathtt{\hat{l}_\mu} = \mathtt{\bar{u}(p_{l^-}) \gamma_\mu (1-\gamma_5)  
v(p_{\nu})~, } 
\end{eqnarray*} 
where $\mathtt{p_\nu}$, $\mathtt{p_l}$ are the momenta of the neutrino 
and of the charged lepton. For radiative semileptonic decays,  
Bremsstrahlung off the charged lepton is not included and must be 
added by hand.  In the anomaly functional,  
only terms containing external gauge fields are included.}\\ \\  
\texttt{ 
Select up to six particles (a photon counts as two particles) 
  of the list $\{\pi_+$, $\pi_-$, $\pi_0$, $K_+$, $K_-$, $K_0$,  
$\bar{K}_0$, $\eta_8$, $\gamma$, $W_+$, $W_-\}$. Assign a momentum to  
each particle. In the list {\it ExternalParticles} all particles are 
  supposed to be incoming. Define the scalar products 
sp[p$_i$, p$_j$] that you want to replace by polynomials of kinematic  
variables in the lists {\it replist1} and {\it replist2}. One should 
not use the capital letters F, G, L, N and R in the lists {\it replist1},  
{\it replist2} and {\it ExternalParticles} as they are 
reserved for coupling constants. Small l is reserved for the 
leptonic matrix element. For semileptonic decays the index $\mu$ 
should only be used for momenta that are contracted with the leptonic 
matrix elements $\mathtt{l_\mu}$ or $\mathtt{\hat{l}_\mu}$. 
Isospin conservation is assumed.}\\ \\  
\texttt{ 
The program terminates if either} \\ 
\texttt{ 
\mbox{}$\;\;\;\;\:$ i) some of the external 
particles are not contained in the list above, \\ 
\mbox{}$\;\;\:$ ii) charge is not 
conserved, \\  
\mbox{}$\:\:$ iii) there are more than one  
external W, \\ 
\mbox{}$\;\;\:$ iv) strangeness change  
$\mathtt{|\Delta S|}$>1. 
} \\ \\ 
\texttt{ 
\noindent  
Notation: \\[.1cm]   
$\mathtt{\bar{A}[M^2]}$ is the renormalized single-propagator loop integral 
$\mathtt{\bar{A}[M^2]=-\displaystyle\frac{M^2}{(4 \pi)^2}\,log(M^2/\mu^2)}$,  
where M is a meson mass and $\mu$ is the renormalization scale.  
$\mathtt{\bar{B}}[s, M_1^2, M_2^2]$ is the standard two-propagator loop 
function subtracted at s=0,  
i.e. $\mathtt{\bar{B}}[0, M_1^2, M_2^2]$=0.  
All loop functions and recursion relations can be found 
in App. B. G$_8$, G$_{27}$ are the nonleptonic weak LECs of order $p^2$.  
$\mathtt{L_i^r, N_i^r,R_i^r}$ are the strong, weak octet and 27-plet LECs 
of O($p^4$), respectively, all renormalized at scale $\mu$. The 
corresponding Lagrangians are reproduced in App. C. 
$\mathtt{\hat{G}_8, \hat{G}_{27}, \hat{N}_i^r, \hat{R}_i^r}$, are complex conjugate LECs.  
\\ \\ 
\underline{Options:} \\ \\ 
Set the following variables equal to 1 to get partial results \\ 
(otherwise the variables should be set to 0): \\ \\ 
onlytreep2=1 $\raum\raum\;\;\;\;$ amplitude of O(p$^2$) \\ 
onlytreep2and4=1 $\raum\;\:$ tree level amplitude up to O(p$^4$) \\ 
onlyloops=1 $\raum\raum\;\;\;\;\;\:$ one-loop amplitude only} \\ \\ 
\texttt{ 
By default (nogmo=0), the Gell-Mann-Okubo relation is applied in the 
amplitudes of O(p$^4$) to express $M_\eta^2$ in terms of $M_K^2$, 
$M_\pi^2$. For better readability, $M_\eta^2$ is 
never replaced  as argument in loop functions. 
Setting nogmo=1, $M_\eta^2$ is kept everywhere.\\
If the recursion relations for the loop integrals should not be 
performed \\
set norecurs=1.} 
 
\begin{verbatim} 
onlytreep2 = 0; 
onlytreep2and4 = 0; 
onlyloops = 0; 
nogmo = 0; 
norecurs=0; 
\end{verbatim} 
\texttt{ 
\noindent (* Example: *) \\  
ExternalParticles = 
$\{\pi_0$[$p_1$],$\pi_0$[$p_2$],$\pi_+$[$p_3$],$\pi_-$[$p_4$]$\}$;\\[.2cm]   
(* scattering amplitude for $\pi_0$[$p_1$] + $\pi_0$[$p_2$] $\to$  
~$\pi_-$[$-p_3$] + $\pi_+$[$-p_4$]  *)\\[.2cm] 
replist1 =$\{$sp[$p_1$,$p_2$],sp[$p_1$,$p_3$],sp[$p_1$,$p_4$], 
sp[$p_2$,$p_3$],sp[$p_2$,$p_4$],sp[$p_3$,$p_4$]$\}$; \\[.2cm] 
replist2 =$\{\displaystyle\frac{1}{2} \left(s -2 M_\pi^2\right)$, 
$\displaystyle\frac{1}{2} \left(t -2 M_\pi^2\right)$, 
$\displaystyle\frac{1}{2} \left(u -2 M_\pi^2\right)$, 
$\displaystyle\frac{1}{2} \left(u -2 M_\pi^2\right)$, 
$\displaystyle\frac{1}{2} \left(t -2 M_\pi^2\right)$,\\[.1cm]  
\hspace*{2cm} $\displaystyle\frac{1}{2} \left(s -2 M_\pi^2\right)\}$;  
\\[.2cm] 
(*  *) \\[.2cm] 
(* Insert the correct path in the next line.*) 
} 
 
\begin{verbatim} 
notesub=NotebookOpen["C:\Math5\programs\Unterdorfer\ampcalculatorsub.nb"];  
 
SelectionMove[notesub,All,Notebook] 
 
SelectionEvaluate[Notebook] 
\end{verbatim} 
 
\setcounter{equation}{0} 
\addtocounter{zaehler}{1} 
\section{Loop integrals and constituent functions} 
\label{app:loops} 
The standard functions $A$, $B$, $B_{ij}$, $C$ and $C_{ij}$ occurring  
in one-loop integrals with up to three propagators are defined through  
the following relations  ($C_F$ denotes the Feynman integration 
contour):  
\glan A \doteq \frac{1}{i}\int_{C_F}\ddk\frac{1}{(k^2-M_P^2)} 
~, \label{eq:A} \glaus 
\glan \{X\} \doteq \frac{1}{i}\int_{C_F}\ddk 
\frac{X}{(k^2-M_P^2)({(k-p)}^2-M_Q^2)} 
~,\glaus 
\glan B(p^2) \doteq \{ 1\} 
~,\glaus 
\glan \{k_\mu\}=p_\mu B_{11}(p^2) 
~,\glaus 
\glan \{k_\mu k_\nu\}=g_{\mu\nu}B_{20}(p^2)+p_\mu p_\nu B_{22}(p^2) 
~,\glaus 
\glan \{\{X\}\} \doteq \frac{1}{i}\int_{C_F}\ddk 
\frac{X}{(k^2-M_P^2)({(k-p_a)}^2-M_Q^2)({(k-p_b)}^2-M_R^2)} 
~,\glaus 
\glan C \doteq \{\{1\}\}~, 
\glaus 
\glan \{\{k_\mu\}\}=p_{a\mu}\,C_{11a}+p_{b\mu}\,C_{11b} 
~,\glaus 
\glan 
\{\{k_\mu k_\nu\}\}=g_{\mu\nu}\,C_{20}+p_{a\mu}p_{a\nu}\,C_{22a}+ 
p_{b\mu}p_{b\nu}\,C_{22b}+(p_{a\mu}p_{b\nu}+p_{b\mu}p_{a\nu}) 
\,\tilde{C}_{22} 
~,\glaus 
\glanf \{\{k_\mu k_\nu k_\lambda \}\} &=& p_{a\mu}p_{a\nu} 
p_{a\lambda}\,C_{33a}+ 
p_{b\mu}p_{b\nu}p_{b\lambda}\,C_{33b}+ 
\nonumber \\ 
& & +(p_{a\mu}p_{a\nu}p_{b\lambda}+p_{a\mu}p_{b\nu}p_{a\lambda} 
+p_{b\mu}p_{a\nu}p_{a\lambda})\,\tilde{C}_{33a}+ 
\nonumber \\ 
& & +(p_{b\mu}p_{b\nu}p_{a\lambda}+p_{b\mu}p_{a\nu}p_{b\lambda} 
+p_{a\mu}p_{b\nu}p_{b\lambda})\,\tilde{C}_{33b}+ 
\nonumber \\ 
& & +(p_{a\mu}g_{\nu\lambda}+p_{a\nu}g_{\mu\lambda}+ 
p_{a\lambda}g_{\mu\nu})\,C_{31a}+ 
\nonumber \\ 
& & +(p_{b\mu}g_{\nu\lambda}+p_{b\nu}g_{\mu\lambda}+ 
p_{b\lambda}g_{\mu\nu})\,C_{31b} 
~,\glausf  \\ 
with $C_{ij\,a}=C_{ij}(p_a^2, p_b^2, p_a\cdot p_b, M^2_P,M^2_Q,M^2_R )$,  
$C_{ij\,b}=C_{ij}(p_b^2, p_a^2, p_a\cdot p_b,M^2_P,M^2_R,M^2_Q )$ 
and $C_{ij}=C_{ij\,a}=C_{ij\,b}$. The  functions $A$ and $B$ can be  
split into a finite and a divergent part for $d \to 4$:
\glan A(M^2_P)=\bar{A}(M^2_P)-2M^2_P \,\Lambda(\mu) \, ,  
\raum B(p^2)=\bar{B}(p^2)+B(0) ~ , 
\glaus 
where 
\glan 
\bar{A}(M^2_P)=-\frac{M_P^2}{(4\pi)^2}\,\mbox{ln}\,\frac{M_P^2}{\mu^2}  
\, ,  \:\:\:\: \Lambda (\mu)=\frac{\mu^{d-4}}{(4\pi)^2} 
\gk\frac{1}{d-4}-\frac{1}{2}(\mbox{ln}(4\pi)+\Gamma'(1)+1)\gK 
~, \label{eq:Lambda} 
\glaus 
\glanf  
\bar{B}(p^2)&=& \frac{1}{32 \pi^2} \,\left( 2+ 
\frac{M^2_P-M^2_Q}{p^2}\, 
\mbox{ln}\, \frac{M^2_Q}{M^2_P} -\frac{M^2_P+M^2_Q}{M^2_P-M^2_Q}\, 
\mbox{ln}\, \frac{M^2_Q}{M^2_P}  - 
\frac{ \sqrt{ \lambda(p^2, M^2_P, M^2_Q)} }{p^2} \right. 
\nonumber \\ && \left. \times\, 
\mbox{ln}\frac{(p^2+ \sqrt{\lambda(p^2, M^2_P, M^2_Q)})^2-(M^2_P-M^2_Q)^2} 
{(p^2- \sqrt{\lambda(p^2, M^2_P, M^2_Q)})^2-(M^2_P-M^2_Q)^2}  \right) \, , 
\nonumber 
\glausf 
\glan 
\lambda(x,y,z)=x^2+y^2+z^2-2(x y+y z+ z x) \, , 
\glaus 
and 
\glan B(0)=-2\Lambda(\mu)+ 
\displaystyle\frac{\bar{A}(M^2_P)-\bar{A}(M^2_Q)}{M^2_P-M^2_Q} \, . 
\label{eq:B0} 
\glaus 
The coefficient functions are of the form 
$$B_{11}(p^2)=\frac{-A({{M_P}}^2) + A({{M_Q}}^2) +  
    B(p^2)\,\left( {{M_P}}^2 - {{M_Q}}^2 + {{p}}^2 \right)  
    }{2\,{{p}}^2}\, , 
$$ 
$$ 
{B_{20}}(p^2) = - \displaystyle\frac{{{p}}^2-3\,{{M_P}}^2 - 3\,{{M_Q}}^2} 
     {288\,{\pi }^2} + \frac{A({{M_Q}}^2) + 2\,B(p^2)\,{{M_P}}^2   
     - \left( {{M_P}}^2 - {{M_Q}}^2 +  
       {{p}}^2 \right) \,{B_{11}}(p^2)}{6} 
\, , $$ 
\glan 
{B_{22}}(p^2) = \displaystyle\frac{{{p}}^2-3\,{{M_P}}^2 - 3\,{{M_Q}}^2} 
     {288\,{\pi}^2 p^2} 
+ \frac{A({{M_Q}}^2) - B(p^2)\,{{M_P}}^2 +  
     2\, \left( {{M_P}}^2 - {{M_Q}}^2 + {{p}}^2 \right) \, 
     {B_{11}}(p^2)}{3\,{{p}}^2} 
\, . \glaus \\ \\ 
The pure $C$ function is given by  
\glanf 
C(p^2_a, p^2_b, p_a\cdot p_b, M^2_P, M^2_Q, M^2_R)=-\frac{1}{16 \pi^2} 
\frac{1}{\sqrt{\lambda}}\sum^3_{i=1}\sum_{\sigma =\pm} 
\left( \mbox{Li}_2\left(\frac{x_i}{x_i-z^\sigma_i}\right) - 
\mbox{Li}_2\left(\frac{x_i-1}{x_i-z^\sigma_i}\right)\right) 
\nonumber \\ 
\glausf 
with the dilogarithm 
\glan \mbox{Li}_2(z)=-\int_0^z \frac{dt}{t}\mbox{ln}\,(1-t) 
\glaus 
and 
\glanf x_1 &=& \frac{1}{2}\left(1+\frac{p^2_a-p^2_b-p^2_D} 
{\sqrt{\lambda}}\right) 
+\frac{1}{2 p^2_a}\left(-M^2_P\left(1+\frac{p^2_a+p^2_D-p^2_b} 
{\sqrt{\lambda}}\right) \right. 
\nl 
&& \left. +M^2_Q\left(1-\frac{p^2_a-p^2_D+p^2_b}{\sqrt{\lambda}} 
\right)\right) 
+\frac{M^2_R}{\sqrt{\lambda}}\, , \nonumber 
\glausf 
\glanf z^\pm_1=\frac{1}{2}\left( 1-\frac{M^2_P-M^2_Q}{p^2_a} \pm 
\sqrt{\left( 1-\frac{M^2_P-M^2_Q}{p^2_a} \right)^2 
-\frac{4}{p^2_a}({M^2_Q}-\mbox{i}\,\epsilon)}\right) \, , \nonumber 
\glausf 
\glanf p^2_D=p^2_a-2\,p_a\cdot p_b+p^2_b\, , \raum 
\lambda=\lambda(p^2_a, p^2_b, p^2_D) \, . 
\glausf 
The quantities $x_2$, $z_2^\sigma$ and $x_3$, $z_3^\sigma$ are obtained by
cyclic interchanges of $(p_a^2, p_D^2, p_b^2)$ and $(M^2_P, M^2_Q, M^2_R)$.

In the following $B(1,3)$ is defined as $B(p_b^2,M^2_P,M^2_R)$ 
and $B(2,3)$ as \\ $B(p_a^2-2\,\qqo+p_b^2,M^2_Q,M^2_R)$. The same 
notation is valid 
for $B_{11}$, $B_{20}$ and $B_{22}$.  $$f_1=p_a^2+M_P^2-M_Q^2\, ,$$ 
$$ {C_{11\,a}} = \frac{  
     {H_{11\,a}}\,{{p_b}}^2- {H_{11\,b}}\,\Mvariable{\qqo}} 
{{{p_a}}^2\,{{p_b}}^2-{\Mvariable{p_a\cdot p_b}}^2} 
\, , $$ $$ 
{C_{20}} = \displaystyle\frac{1}{64\,{\pi }^2} +  
\frac{2\,C\,M_P^2 +  B(2,3) -  
     {C_{11\,a}}\,f_1 - {C_{11\,b}}\,({{M_P}}^2 - {{M_R}}^2 + {{p_b}}^2)}{4} 
\, , $$ $$ 
{C_{22\,a}} = \frac{  
     {H_{21\,a}}\,{{p_b}}^2- 
{H_{22\,b}}\,\Mvariable{\qqo}}{{{p_a}}^2\, 
{{p_b}}^2-{\Mvariable{\qqo}}^2} 
\, , \raum {{\tilde{C}}_{22}} = \frac{  
     {H_{22\,b}}\,{{p_a}}^2- {H_{21\,a}}\, \Mvariable{\qqo}} 
{{{p_a}}^2\,{{p_b}}^2-{\Mvariable{\qqo}}^2} 
\, , $$ $$ 
{C_{33\,a}} = \frac{  
     {H_{31\,a}}\,{{p_b}}^2-{H_{32\,b}} \,\Mvariable{\qqo}} 
{{{p_a}}^2\,{{p_b}}^2-{\Mvariable{\qqo}}^2 } 
\, , \raum {{{\tilde{C}}}_{33\,a}} = \frac{  
     {H_{32\,b}}\,{{p_a}}^2- {H_{31\,a}} \,\Mvariable{\qqo}} 
{{{p_a}}^2\,{{p_b}}^2-{\Mvariable{\qqo}}^2} 
\, , $$ $$ 
{C_{31\,a}} = \frac{  
     {H_{30\,a}}\,{{p_b}}^2-{H_{30\,b}} \,\Mvariable{\qqo}} 
{{{p_a}}^2\,{{p_b}}^2-{\Mvariable{\qqo}}^2} 
\, , $$ \\  $$ 
{H_{11\,a}} = \frac{-B(1,3) + B(2,3) + C\,{f_1}}{2} 
\, , \raum {H_{21\,a}} = \frac{B(2,3) + {C_{11\,a}}\,{f_1} -  
{B_{11}}(2,3)}{2}-C_{20} 
\, , $$ $$ 
{H_{22\,a}} =\frac{{C_{11\,b}}\,{f_1} - {B_{11}}(1,3) + {B_{11}}(2,3)}{2} 
\, , \raum {H_{30\,a}} = \frac{{C_{20}}\,{f_1} - {B_{20}}(1,3) +  
{B_{20}}(2,3)}{2} 
\, , $$ $$ 
{H_{31\,a}} = \frac{B(2,3) + {f_1}\,{{{{C}}}_{22\,a}} -  
    2\,{B_{11}}(2,3) + {B_{22}}(2,3)}{2}-2C_{31a} 
\, , $$ $$ \raum {H_{32\,a}} = \frac{{f_1}\,{{{C}}_{22\,b}} -  
{B_{22}}(1,3) +  
    {B_{22}}(2,3)}{2} 
\, . $$ 
\glan \glaus 
The results for the integrals agree with those in  
Ref.~\cite{Passarino:1978jh}. 
 
The constituent functions of the one-loop functional are of the  
following form: 
\glanf & & G_{1\,PQ}^{\mu\nu}(p)=p^\mu p^\nu \, a(p^2,M^2_P,M^2_Q) 
+ g^{\mu\nu} \, b(p^2,M^2_P,M^2_Q)\, , 
\nonumber \\ \nonumber \\ 
& & G_{2\,PQ}^{\mu}(p)=i\, p^\mu\,\left(\frac{1}{2}\,B(p^2,M^2_P,M^2_Q) 
-B_{11}(p^2,M^2_P,M^2_Q)\right) \, ,\raum 
\nonumber \\ \nonumber \\ 
& & G_{3\,PQ}(p)=\frac{1}{4} B(p^2,M^2_P,M^2_Q) \, ,\nonumber \\  
\nonumber \\ 
& & G_{4\,PQR}^{\mu\nu\lambda}(p_a,p_b)=i\,\gk c\aaa\,\ppm\ppn\ppl  
+c\bbb\,\PPm\PPn\PPl 
+d\aaa\,\ppm\ppn\PPl 
\nonumber \\ 
& & \raum +e\aaa\,\ppm\PPn\ppl+f\aaa\,\PPm\ppn\ppl 
+d\bbb\,\PPm\PPn\ppl+f\bbb\,\PPm\ppn\PPl  
\nonumber \\ 
& & \raum  +e\bbb\,\ppm\PPn\PPl 
+g\aaa\,\ppm g^{\nu\lambda}+h\aaa\,\ppn g^{\mu\lambda}+ 
i\aaa\,\ppl g^{\mu\nu}  
\nonumber \\ 
& & \raum +h\bbb\,\PPm g^{\nu\lambda}+g\bbb\,\PPn g^{\mu\lambda} + 
i\bbb\,\PPl g^{\mu\nu}\gK \, , 
\nonumber \\ \nonumber \\ 
& & G_{5\,PQR}^{\mu\nu}(p_a,p_b)=j\aaa\, g^{\mu\nu} +k\aaa\,\ppm\ppn+ 
k\bbb\,\PPm\PPn 
\nonumber \\ 
& & \raum  +l\aaa\,\ppm\PPn+m\aaa\,\PPm\ppn \, , 
\nonumber \\ \nonumber \\ 
& & G_{6\,PQR}^{\mu}(p_a,p_b)=i\,\gk n\aaa\,\ppm + 
n\bbb\,\PPm \gK\, , \raum 
G_{7\,PQR}(p_a,p_b)=\frac{1}{6}\,C ~, 
\glausf 
with for example 
\glan c_a=c(p_a^2, p_b^2, p_a\cdot p_b,M^2_P,M^2_Q,M^2_R)\raum\mbox{and} 
\raum c_b=c(p_b^2, p_a^2, p_a\cdot p_b,M^2_P,M^2_R,M^2_Q) \, , 
\glaus 
where 
\glanf 
&&a(p^2,M^2_P,M^2_Q )= 
\frac{ 4({B}_{11}(p^2,M^2_P,M^2_Q )-{B_{22}}(p^2,M^2_P,M^2_Q ))-  
{B}(p^2,M^2_P,M^2_Q )} 
  {4}  \, , \nonumber \\ \nonumber \\ 
&&b(p^2,M^2_P,M^2_Q)= 
-{B_{20}}(p^2,M^2_P,M^2_Q ) \, ,\nonumber \\ \nonumber \\ 
&&c_a= 
\frac{{C_{11\,a}}  -  
    4\,C_{22\,a}+ 4\,{C_{33\,a}}}{3} \, , \raum 
d_a= 
\frac{ {C_{11\,a}} -  
    2\,C_{22\,a} -2\,{{\tilde{C}}_{22}}  +  
    4\,{{\tilde{C}}}_{33\,a}}{3} \, ,\nonumber \\ \nonumber \\ 
&& e_a= 
\frac{ -C/2+2\,{C_{11\,a}} +  
    {C_{11\,b}} -  
    2\,C_{22\,a}-4\,{{\tilde{C}}_{22}} +  
    4\,{{\tilde{C}}}_{33\,a} }{3}  \, ,\nonumber \\ \nonumber \\ 
&& f_a= 
\frac{-2\,\left( {{\tilde{C}}_{22}} -  
      2\,{{\tilde{C}}}_{33\,a} \right) }{3}  \, , \raum 
g_a= 
\frac{-2\,\left( {C_{20}} - 2\,{C_{31\,a}} \right) }{3} \, , \raum 
h_a= 
\frac{4\,{C_{31\,a}}}{3} \, ,\nonumber \\ \nonumber \\ 
&& i_a= 
\frac{-2\,\left( {C_{20}} - 2\,{C_{31\,a}} \right) }{3} \, , \raum 
j_a= 
-2\,{C_{20}} \, , \raum 
k_a= 
{C_{11\,a}} - 2\,C_{22\,a} \, , \nonumber \\ \nonumber \\ 
&& l_a= 
- \frac{{C}}{2} + {C_{11\,a}} + {C_{11\,b}}-2\,{{\tilde{C}}_{22}}  
\, , \raum 
m_a= 
-2\,{{\tilde{C}}_{22}} \, , \raum 
n_a= 
\frac{ {C}}{2}-2\,{C_{11\,a}} ~ . \nonumber \\ 
\end{eqnarray}  
 
\setcounter{equation}{0} 
\addtocounter{zaehler}{1} 
\section{Lagrangians of $\mbf{O(p^4)}$} 
\label{app:lagr4} 
The strong Lagrangian of $O(p^4)$ is given by \cite{Gasser:1984gg} 
\glanf 
{\cal L}^S_4 & = & L_1 \langle D_\mu U^\dg D^\mu U\rangle^2 + 
                 L_2 \langle D_\mu U^\dg D_\nu U\rangle 
                     \langle D^\mu U^\dg D^\nu U\rangle \nl 
& & + L_3 \langle D_\mu U^\dg D^\mu U D_\nu U^\dg D^\nu U\rangle + 
    L_4 \langle D_\mu U^\dg D^\mu U\rangle \langle \chi^\dg U + 
    \chi U^\dg\rangle  \nl 
& & +L_5 \langle D_\mu U^\dg D^\mu U(\chi^\dg U + U^\dg 
\chi)\rangle 
    + 
    L_6 \langle \chi^\dg U + \chi U^\dg \rangle^2  \nl
& &  + 
    L_7 \langle \chi^\dg U - \chi U^\dg \rangle^2 
+ L_8 \langle \chi^\dg U \chi^\dg U + 
 \chi U^\dg \chi U^\dg\rangle 
    \nl 
& &  -i L_9 \langle F_R^{\mu\nu} D_\mu U D_\nu U^\dg + 
      F_L^{\mu\nu} D_\mu U^\dg D_\nu U \rangle 
+ L_{10} \langle U^\dg F_R^{\mu\nu} U F_{L\mu\nu}\rangle  
\label{eq:Lag4S} 
\glausf 
with 
\glan F_{\mu\nu}^R=\dmuu r_\nu -\dnuu r_\mu -\ii\,[r_\mu, r_\nu]\, , \raum 
F_{\mu\nu}^L=\dmuu l_\nu -\dnuu l_\mu -\ii\,[l_\mu, l_\nu] ~. 
\glaus 
 
The nonleptonic weak Lagrangian of $O(G_F p^4)$ consists of an octet 
and a 27-plet part: 
\begin{equation}  
\cL_4^W = \cL_4^{W,8} + \cL_4^{W,27} ~. 
\label{eq:L4W} 
\end{equation} 
Neglecting terms that would only contribute at $O(G_F^2)$ (with 
external $W^\pm$ fields in addition to the nonleptonic weak transition),  
the $\Delta S=1$ octet Lagrangian can be written as  
\cite{Kambor:1989tz,Ecker:1992de} 
\glanf 
\cL_4^{W,8} 
&=& G_8 F^2 \{ 
N_1\; \langle \lambda D_\mu U^\dg D^\mu U D_\nu U^\dg D^\nu U \rangle 
+ N_2\; \langle \lambda D_\mu U^\dg D^\nu U D_\nu U^\dg D^\mu U 
\rangle \nl 
&& + N_3\; \langle \lambda D_\mu U^\dg D_\nu U\rangle \langle D^\mu U^\dg D^\nu 
U \rangle + 
N_4\; \langle \lambda D_\mu U^\dg U\rangle \langle U^\dg D^\mu U D_\nu 
U^\dg D^\nu U \rangle \nl 
&& + N_5\; \langle \lambda \{ S, D_\mu U^\dg D^\mu U\} 
\rangle +  
N_6\; \langle \lambda D_\mu U^\dg U\rangle \langle U^\dg D^\mu U 
S\rangle \nl 
&& + N_7\; \langle \lambda S\rangle \langle D_\mu U^\dg 
D^\mu U \rangle + 
N_8\; \langle \lambda D_\mu U^\dg D^\mu U\rangle \langle S \rangle \nl 
&& +  N_9\;i  \langle \lambda [P, D_\mu U^\dg D^\mu U] 
\rangle + 
N_{10}\; \langle \lambda S^2\rangle \nl 
&& + N_{11}\; \langle \lambda S\rangle \langle S \rangle - 
N_{12}\; \langle \lambda P^2\rangle  
- N_{13}\; \langle \lambda P \rangle \langle P \rangle \label{eq:L4W8}\\ 
&& + 
N_{14}\; i \langle \lambda \{ V^{\mu\nu}, 
D_\mu U^\dg D_\nu U\} \rangle  
 + N_{15}\; i \langle \lambda D_\mu U^\dg U V^{\mu\nu} U^\dg  
D_\nu U \rangle \nl && - 
N_{16}\; i \langle \lambda \{ A^{\mu\nu}, 
D_\mu U^\dg D_\nu U\} \rangle  
 - N_{17}\; i \langle \lambda D_\mu U^\dg U A^{\mu\nu} U^\dg 
D_\nu U \rangle \nl && + 
2 N_{18}\; \langle \lambda (F_L^{\mu\nu} U^\dg F_{R\mu\nu} U + 
U^\dg F_{R\mu\nu} U F_L^{\mu\nu}) \rangle \nl 
&& + N_{28}\; i \ve_{\mu\nu\rho\sigma} \langle \lambda D^\mu U^\dg U \rangle 
\langle U^\dg D^\nu U D^\rho U^\dg D^\sigma U \rangle \nl &&+ 
2 N_{29}\; \ve_{\mu\nu\rho\sigma} \langle \lambda [U^\dg  
F_R^{\mu\nu} U, D^\rho U^\dg D^\sigma U] \rangle \nl 
&& + N_{30}\; \ve_{\mu\nu\rho\sigma} \langle \lambda  
U^\dg D^\mu U \rangle \langle V^{\rho\sigma} D^\nu U^\dg U \rangle \nl 
&& - N_{31}\; \ve_{\mu\nu\rho\sigma} \langle \lambda  
U^\dg D^\mu U \rangle \langle A^{\rho\sigma} D^\nu U^\dg U \rangle 
 \} 
+ {\rm h.c.}  \nonumber 
\glausf 
with 
\glanf 
S&=&\chi^\dag U+U^\dag\chi\, , \raum 
P=\ii\, (\chi^\dag U-U^\dag\chi) \, ,\nl 
V_{\mu\nu}&=&U^\dag F^R_{\mu\nu} U+F^L_{\mu\nu} \, , \raum 
A_{\mu\nu}=U^\dag F^R_{\mu\nu} U-F^L_{\mu\nu} ~ . 
\glausf 
 
For the 27-plet Lagrangian we use the form given in  
Ref.~\cite{Esposito-Farese:1990yq}: 
\glan 
\cL_4^{W,27} = G_{27} F^2 \tilde{t}_{ij,kl} I_{ji,lk} 
+ {\rm h.c.} ~ , \label{eq:L4W27} 
\glaus 
\glanf 
I_{ij,kl}&=&{R_1}\,{{\left( L^{\mu }\,{L_{\mu }} \right) }_ 
     {\Mvariable{ij}}}\, 
   {{\left( L^{\nu }\,{L_{\nu }} \right) }_ 
     {\Mvariable{kl}}} +  
  {R_2}\,{{\left( {L_{\mu }}\,{L_{\nu }} \right) }_ 
     {\Mvariable{ij}}}\, 
   {{\left( {L^{\mu }}\,{L^{\nu }} \right) }_ 
     {\Mvariable{kl}}} \nl 
 && +  
  {R_3}\,{{\left( {L_{\mu }}\,{L_{\nu }} \right) }_ 
     {\Mvariable{ij}}}\, 
   {{\left( {L^{\nu }}\,{L^{\mu }} \right) }_ 
     {\Mvariable{kl}}} +  
  {R_4}\,\left(L_{\mu }\right)_{\Mvariable{ij}} \, 
   \left( L_{\nu }\,L^{\mu }\,L^{\nu } \right)_{\Mvariable{kl}} \nl 
&&  +  
  {R_5}\,\left(L_{\mu }\right)_{\Mvariable{ij}}\, 
   {{\{ {L^{\mu }},\,{{L_{\nu }}}\,{{L^{\nu }}} \} }_{\Mvariable{kl}}}  
+  
  {R_6} \, \kk L^{\mu}\, L_{\mu} \KK \,  \left( L_{\nu}\right)_{\Mvariable{ij}}\, 
   \left( L^{\nu }\right)_{\Mvariable{kl}}  \nl 
&& +   
  {R_7}\,{S_{\Mvariable{ij}}}\,{{\left( L^{\mu }\,{L_{\mu }} \right) }_ 
     {\Mvariable{kl}}} 
+  
 {R_8}\,  
   {{\{ S,\,{L_{\mu }} \} }_{\Mvariable{ij}}} 
{{\left( L^{\mu } \right) }_{\Mvariable{kl}}}\,  \nl 
&& +  
 {R_9}\,\ii\, 
   {{[ P,\,{L_{\mu }} ] }_{\Mvariable{ij}}} 
{{\left( L^{\mu } \right) }_{\Mvariable{kl}}}\,  
+ 
  {R_{10}}\, \kk S \KK\,\left(L_{\mu }\right)_{\Mvariable{ij}}\, 
{{\left( L^{\mu } \right) }_{\Mvariable{kl}}}\,  
    \nl 
&& +  
  {R_{11}}\,{S_{\Mvariable{ij}}}\,{S_{\Mvariable{kl}}} 
+ 
{R_{12}}\,{P_{\Mvariable{ij}}}\,{P_{\Mvariable{kl}}}  
 +  {R_{13}}\,\ii \,\left(v_{\Mvariable{\mu \nu }}\right)_{\Mvariable{ij}}  
\,{{\left( {L^{\mu }}\,{L^{\nu }} \right) }_ 
     {\Mvariable{kl}}}  \\ 
&& + R_{14}\,\ii\,\left(a_{\mu\nu}\right)_{ij}\left(L^\mu L^\nu \right)_{kl} 
+R_{15}\,\ii\,\left[v_{\mu\nu},\, L^\mu\right]_{ij}\, 
 \left(L^{\nu }\right)_{\Mvariable{kl}} \nl 
&&+R_{16}\,\ii\,\left[a_{\mu\nu},\, L^\mu\right]_{ij}\, 
 \left(L^{\nu }\right)_{\Mvariable{kl}} 
+R_{17} \,\left(v_{\mu\nu}\right)_{ij}\,\left(v^{\mu\nu}\right)_{kl} \nl 
&& +R_{18} \,\left(v_{\mu\nu}\right)_{ij}\,\left(a^{\mu\nu}\right)_{kl} 
+R_{19} \,\left(a_{\mu\nu}\right)_{ij}\,\left(a^{\mu\nu}\right)_{kl} \nl 
&& +R_{20}\,\ii\,\epsilon^{\mu\nu\rho\sigma}\, 
{{\left( {L_{\mu }}\,{L_{\nu }} \right) }_ 
     {\Mvariable{ij}}}\, 
   {{\left( {L_{\rho }}\,{L_{\sigma }} \right) }_ 
     {\Mvariable{kl}}} 
+R_{21}\,\ii\,\epsilon^{\mu\nu\rho\sigma}\, 
{{\left( {L_{\mu }}  \right) }_ 
     {\Mvariable{ij}}}\, 
   {{\left( {L_{\nu }}\,{L_{\rho }}\,{L_{\sigma }} \right) }_ 
     {\Mvariable{kl}}} \nl 
&& +R_{22}\,\epsilon^{\mu\nu\rho\sigma}\, 
{{\left( {L_{\mu }} \right) }_ 
     {\Mvariable{ij}}}\, 
 {{\left( {L_{\nu }}\,v_{\rho\sigma} \right) }_ 
     {\Mvariable{kl}}} 
+R_{23}\,\epsilon^{\mu\nu\rho\sigma}\, 
{{\left( {L_{\mu }}  \right) }_ 
     {\Mvariable{ij}}}\, 
   {{\left( {L_{\nu }}\,a_{\rho\sigma} \right) }_ 
     {\Mvariable{kl}}} \nonumber ~. 
\glausf 
The coefficients $\tilde{t}_{ij,kl}$ are defined as 
\glanf && \tilde{t}_{12,31}=\tilde{t}_{31,12} 
=\tilde{t}_{32,11}=\tilde{t}_{11,32}= \frac{1}{3} 
\, , \nl 
&& \tilde{t}_{22,32}=\tilde{t}_{32,22}= 
\tilde{t}_{32,33}=\tilde{t}_{33,32}= - \frac{1}{6} \, , \nl  
&& \tilde{t}_{ij,kl}=0 \raum\mbox{otherwise} ~. \label{eq:tijkl2} 
\glausf 
The low-energy constants $L_i$, $N_i$ and $R_i$ are dimensionless. 
Moreover, the $L_i$ are real whereas $G_8$, $G_{27}$, $N_i$, $R_i$  
can be complex in the presence of CP violation.

\newpage 

\end{document}